\documentclass[%
 reprint,
 superscriptaddress,
 showkeys,
 showpacs,
 amsmath,amssymb,
 aps,
]{revtex4-1}

\usepackage{graphicx}
\usepackage{dcolumn}
\usepackage{bm}
\usepackage{todonotes}

\begin{document}


\title{Reconstruction of cosmic ray air showers with Tunka-Rex data using template fitting of radio pulses}

\author{P.A.~Bezyazeekov}
\affiliation{Institute of Applied Physics ISU, Irkutsk, 664020 Russia}

\author{N.M.~Budnev}
\affiliation{Institute of Applied Physics ISU, Irkutsk, 664020 Russia}

\author{D.~Chernykh}
\affiliation{Institute of Applied Physics ISU, Irkutsk, 664020 Russia}

\author{O.~Fedorov}
\affiliation{Institute of Applied Physics ISU, Irkutsk, 664020 Russia}

\author{O.A.~Gress}
\affiliation{Institute of Applied Physics ISU, Irkutsk, 664020 Russia}

\author{A.~Haungs}
\affiliation{Institut f\"ur Kernphysik, Karlsruhe Institute of Technology (KIT), Karlsruhe, 76021 Germany}

\author{R.~Hiller}
\email{now at the University of Z\"urich}
\affiliation{Institut f\"ur Kernphysik, Karlsruhe Institute of Technology (KIT), Karlsruhe, 76021 Germany}

\author{T.~Huege}
\email{also at Vrije Universiteit Brussel, Brussels, Belgium}
\affiliation{Institut f\"ur Kernphysik, Karlsruhe Institute of Technology (KIT), Karlsruhe, 76021 Germany}

\author{Y.~Kazarina}
\affiliation{Institute of Applied Physics ISU, Irkutsk, 664020 Russia}

\author{M.~Kleifges}
\affiliation{Institut f\"ur Prozessdatenverarbeitung und Elektronik, Karlsruhe Institute of Technology (KIT), Karlsruhe, 76021 Germany}

\author{D.~Kostunin}
\email{dmitriy.kostunin@kit.edu}
\affiliation{Institut f\"ur Kernphysik, Karlsruhe Institute of Technology (KIT), Karlsruhe, 76021 Germany}

\author{E.E.~Korosteleva}
\affiliation{Skobeltsyn Institute of Nuclear Physics MSU, Moscow, 119991 Russia}

\author{L.A.~Kuzmichev}
\affiliation{Skobeltsyn Institute of Nuclear Physics MSU, Moscow, 119991 Russia}

\author{V.~Lenok}
\affiliation{Institut f\"ur Kernphysik, Karlsruhe Institute of Technology (KIT), Karlsruhe, 76021 Germany}

\author{N.~Lubsandorzhiev}
\affiliation{Skobeltsyn Institute of Nuclear Physics MSU, Moscow, 119991 Russia}

\author{T.~Marshalkina}
\affiliation{Institute of Applied Physics ISU, Irkutsk, 664020 Russia}

\author{R.R.~Mirgazov}
\affiliation{Institute of Applied Physics ISU, Irkutsk, 664020 Russia}

\author{R.~Monkhoev}
\affiliation{Institute of Applied Physics ISU, Irkutsk, 664020 Russia}

\author{E.~Osipova}
\affiliation{Skobeltsyn Institute of Nuclear Physics MSU, Moscow, 119991 Russia}

\author{A.~Pakhorukov}
\affiliation{Institute of Applied Physics ISU, Irkutsk, 664020 Russia}

\author{L.~Pankov}
\affiliation{Institute of Applied Physics ISU, Irkutsk, 664020 Russia}

\author{V.V.~Prosin}
\affiliation{Skobeltsyn Institute of Nuclear Physics MSU, Moscow, 119991 Russia}

\author{F.G.~Schr\"oder}
\affiliation{Institut f\"ur Experimentelle Teilchenphysik, Karlsruhe Institute of Technology (KIT), Karlsruhe, 76021 Germany}

\author{D.~Shipilov}
\affiliation{Institute of Applied Physics ISU, Irkutsk, 664020 Russia}

\author{A.~Zagorodnikov}
\affiliation{Institute of Applied Physics ISU, Irkutsk, 664020 Russia}

\collaboration{Tunka-Rex Collaboration}

\date{\today}

\begin{abstract}
We present an improved method for the precise reconstruction of
cosmic-ray air showers above $10^{17}$~eV with sparse radio arrays.
The method is based on the comparison of measured pulses to predictions for radio pulse shapes by CoREAS simulations.
We applied our method to the data of Tunka-Rex, a 1~km\textsuperscript{2} radio array in Siberia operating in the frequency band of 30-80~MHz.
Tunka-Rex is triggered by the air-Cherenkov detector Tunka-133 and by scintillators (Tunka-Grande).
The instrument collects air-shower data since 2012.
The present paper describes an updated data analysis of Tunka-Rex and details of the new method applied.
After quality cuts, when Tunka-Rex reaches its full efficiency, the energy resolution of about 10\% given by the new method has reached the limit of systematic uncertainties due to the calibration uncertainty and shower-to-shower fluctuations.
At the same time the shower maximum reconstruction has improved compared to the previous method based on the slope of the lateral distribution and reaches a precision of better than 35~g/cm\textsuperscript{2}.
We also define conditions of the measurements at which the shower maximum resolution of Tunka-Rex reaches a value of 25~g/cm\textsuperscript{2} and becomes competitive to optical detectors.
To check and validate our reconstruction and efficiency cuts we compare individual events to the reconstruction of Tunka-133.
Furthermore, we compare the mean of the shower maximum as a function of primary energy to the measurements of other experiments.
\end{abstract}

\maketitle


\section{Introduction}
The energy spectrum and the mass composition of ultrahigh energy cosmic rays (with primary energies above 100~PeV) is of special interest
since it sheds light on the transition from galactic to extragalactic accelerators.
There are many methodological approaches of decoding energy and composition spectra using particle and optical detectors~\cite{AUGER_TA_ENERGY_ICRC2017, AUGER_TA_MASS_ICRC2017}.
Meanwhile, radio detectors have shown their ability of precise reconstruction of air showers produced by ultra-high energy cosmic rays~\cite{Huege:2016veh,Schroder:2016hrv}.
Tunka-Rex was the first large-scale, sparse radio array which has shown that it is possible to reconstruct the primary energy with a resolution of about 15\% and shower maximum with a competitive resolution of about 40~g/cm\textsuperscript{2}, even having only a few antennas involved per event~\cite{Bezyazeekov:2015ica}.
LOFAR with its very dense layout (hundreds antenna stations per event) achieved a shower maximum resolution comparable with optical detectors~\cite{Buitink:2014eqa}, however it will be very difficult (due to costs) to build large-scale detectors focused on energies higher than $10^{17.5}$~eV with this density.
To complement studies performed in this energy range by optical detectors, radio detectors should feature the same energy and shower maximum resolution (10\% and 20~g/cm\textsuperscript{2} respectively).
This is crucial for the next-generation radio detectors focused on detection of gamma ray photons~\cite{V.:2017kbm} and neutrino~\cite{Martineau-Huynh:2017bpw} .

A step toward high precision is completed in the present approach:
we try to exploit as much information as possible from the radio pulse to perform a precise reconstruction of the primary energy and the depth of the shower maximum.
The existing standard reconstruction by Tunka-Rex uses only the pulse maxima~\cite{Bezyazeekov:2015ica}, and the new method additionally makes use of the pulse shape.
Based on the standard procedure the reconstructed events are reproduced with CoREAS~\cite{Huege:2013vt} simulations for different primary particles.
Then the pulse shapes of simulated radio pulses are fitted to the measured ones.

In this work, we discuss requirements and advantages of the proposed method,
describe the details of the updated Tunka-Rex reconstruction,
and present a cross-check of the Tunka-Rex results with Tunka-133~\cite{Prosin:2016jev}.
As a result the mean of the shower maximum as a function of primary energy is reconstructed from Tunka-Rex events (triggered by Tunka-133) acquired from 2012.

\section{Detector description and calibration}
The Tunka Radio Extension (Tunka-Rex) is a radio array for the detection of cosmic rays in the energy range of $10^{17}$ to $10^{18}$~eV.
It is located at the Tunka Advanced Instrument for cosmic rays and Gamma Astronomy (TAIGA)~\cite{Budnev:2017jyl} near Lake Baikal, Siberia.
The radio array is equipped with 63 antenna stations measuring radio emission in the frequency band of 30-80~MHz, distributed on 1 km\textsuperscript{2}.
Each antenna station consists of two short aperiodic loaded loop antennas (SALLA)~\cite{Abreu:2012pi}, aligned perpendicularly to each other in the horizontal plane.
The signal at the antennas is amplified and digitalized with a sampling rate of 200 MS/s, and recorded in traces with 1024 samples each.
The basic description of Tunka-Rex is given in Ref.~\cite{TunkaRex_NIM_2015} and the details of upgrade and latest results are given in Ref.~\cite{Kostunin:2017bzd}.

To reconstruct the electric field at the antenna it is necessary to know the hardware response of the antenna station, namely antenna pattern, and the gain and phase responses of the electronics.
The signal circuit of Tunka-Rex was calibrated under laboratory conditions.
The antenna pattern and phase response were calculated with the simulation code NEC2~\cite{nec2}, then a calibration of the absolute gain was performed~\cite{TunkaRex_NIM_2015}.
The absolute amplitude calibration of the Tunka-Rex antenna station was performed with the same reference source as for LOPES~\cite{LOPES:2015eya}
which enabled us to perform a cross-check between KASCADE-Grande and Tunka-133 energy scales~\cite{Apel:2016gws}.

\begin{figure}
\includegraphics[width=0.49\linewidth]{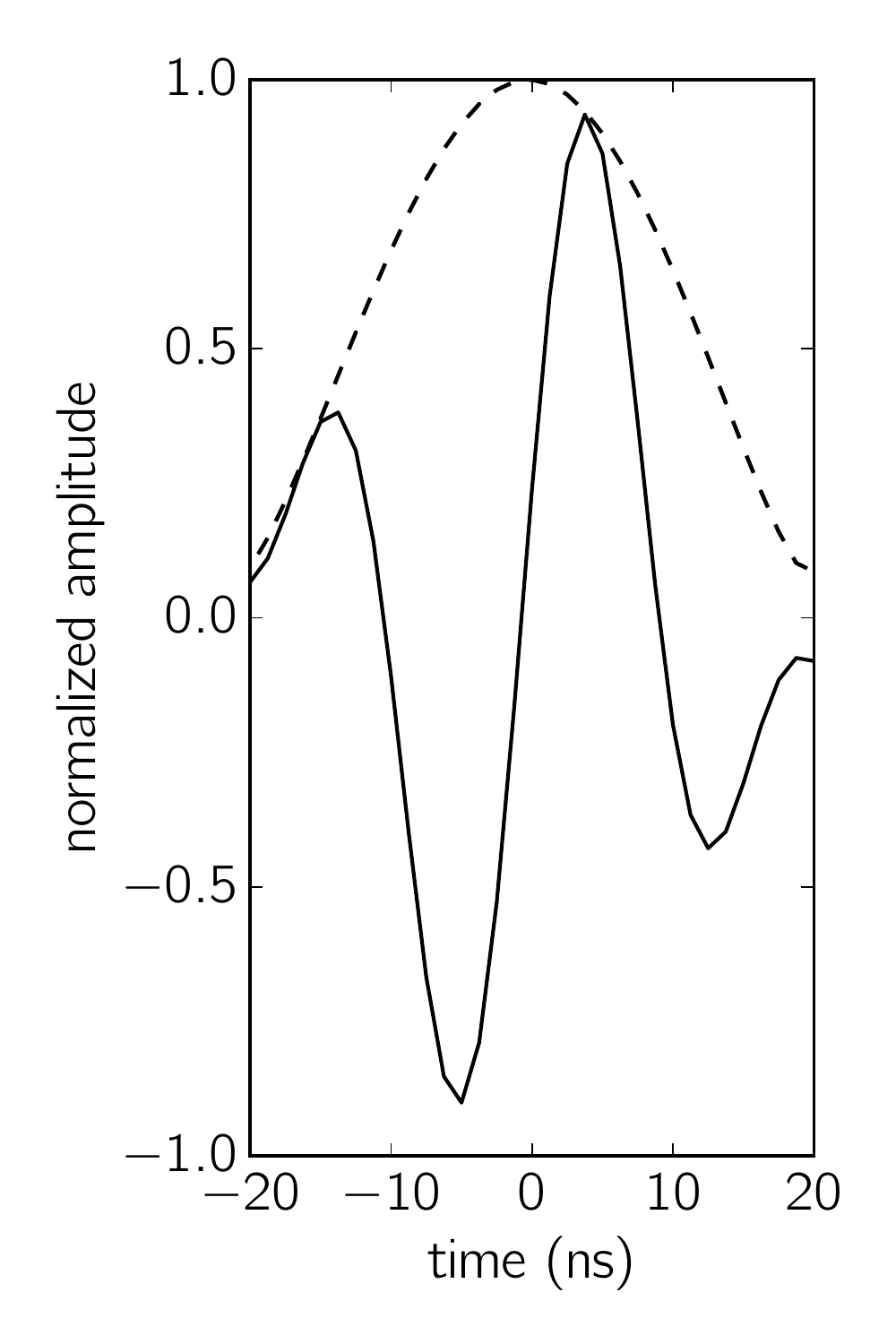}
\includegraphics[width=0.49\linewidth]{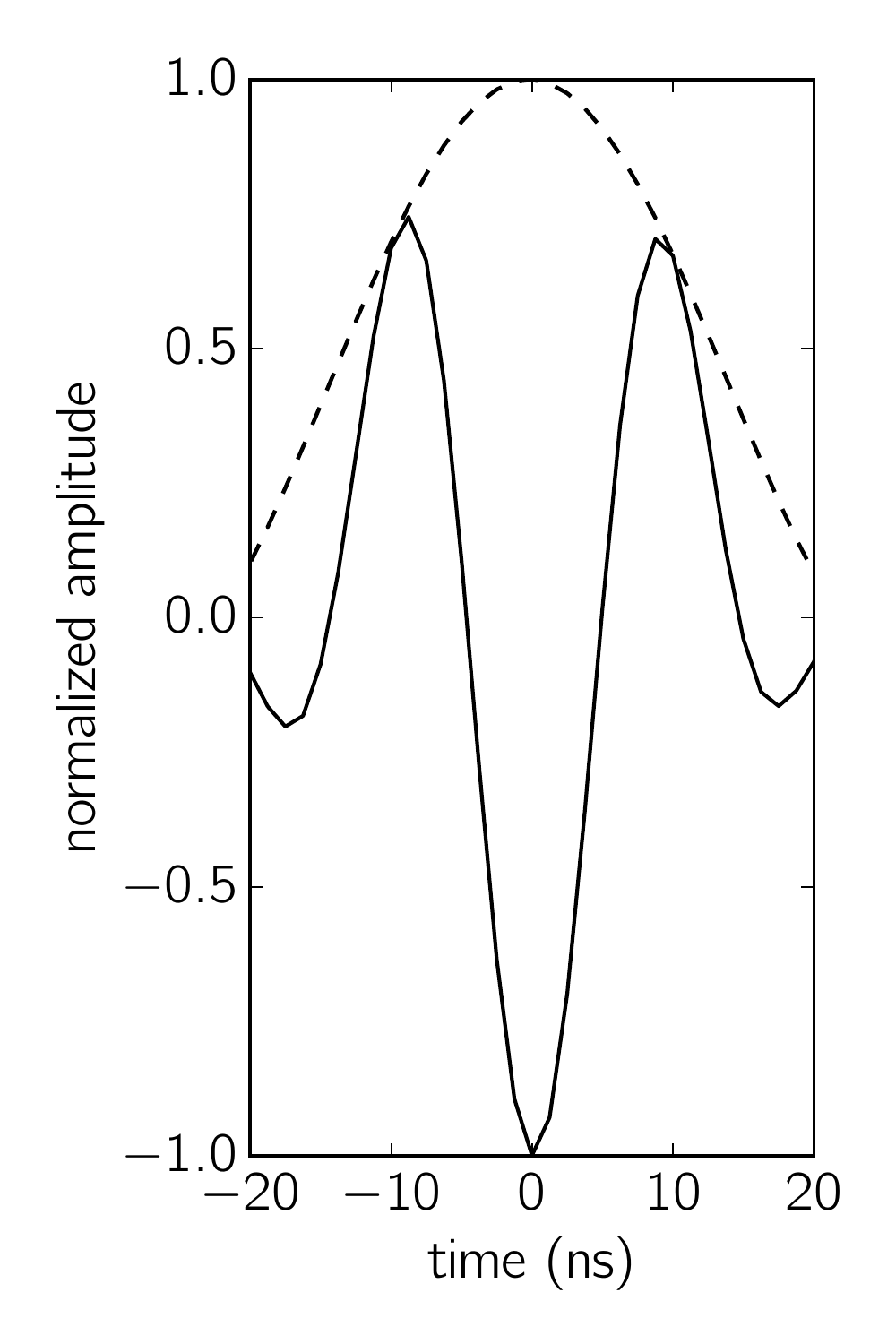}
\caption{Averaged measured (\textit{left}) and simulated (\textit{right}) signals with their envelopes (instantaneous amplitudes).
One can note that the pulse shapes differ significantly from each other (which indicates the difference in phase response), while the shapes of the envelopes are in agreement (which indicates agreement in the gain response).}
\label{fig:signals}
\end{figure}

In Ref.~\cite{Kostunin_TunkaRex_ICRC2017} we suggested an approach for $X_\mathrm{max}$ reconstruction which uses the full information of the radio measurements, i.e. uses measured electric fields at the antennas (instead of only the maximum of signal amplitudes or signal powers).
Upon closer inspection, we have found that our phase calibration does not provide sufficient accuracy for application of this approach.
One can see the difference between simulated and reconstructed pulses in Fig.~\ref{fig:signals}, which would introduce a significant systematic uncertainty in the analysis.
Nevertheless, the envelopes (i.e. instantaneous amplitudes) of the signals are still in very good agreement and used in the analysis described in the present paper.
This means that in contrast to the frequency-dependent phase response, the frequency-dependent gain of all instrumental components is understood sufficiently well.
Using the full shape of the envelope the information content of each measurement is increased by several times compared to prior methods which use single observables per antenna station, such as the maximum amplitude or pulse integral.


\section{Signal reconstruction}
\label{sec:signalrec}
The envelope of the signal $s(t)$ is defined as the absolute value of the analytic signal ${u(t) = s(t) + i\mathcal{H}[s(t)]}$, where $\mathcal{H}$ denotes the Hilbert transformation.
Since we measure two polarization directions of the air-shower signal, the resulting envelope is calculated as follows:
\begin{equation}
u(t) = \sqrt{u_{\mathbf{v}\times \mathbf{B}}^2(t) + u_{\mathbf{v}\times \mathbf{v} \times \mathbf{B}}^2(t)}\,,
\end{equation}
where $u_{\mathbf{v}\times \mathbf{B}}^2(t)$ and $u_{\mathbf{v}\times \mathbf{v} \times \mathbf{B}}^2(t)$ are components of the electric fields measured in the shower plane~\cite{Kostunin:2015taa},
the third component perpendicular to them contributes only to less than 2\%~\cite{Apel:2014usa} and is neglected~\footnote{This component can be included in  the analysis of very-high SNR events, however the expected improvement in the resolution of the shower maximum is estimated to less than 1~g/cm\textsuperscript{2}}.

Compared to our previous analysis we optimized the estimation of the signal-to-noise ratio (SNR) and the signal selection.
Since we occasionally have RFI in the signal and noise windows, especially after upgrades at the TAIGA facility, the following improvements were introduced:

\begin{itemize}
\item \emph{The full width of a pulse} is limited to 50~ns.
Hereafter we define the pulse width as the distance between the two minima of the envelope closest to the peak (in Fig.~\ref{fig:signals} the peak of the amplitude is at 0~ns, and the closest minima are at $-20$ and $20$~ns, i.e. the full width of the pulse is $40$~ns).
To prevent low-frequency RFI passing through the signal window all ``broad'' pulses (with a width of more than 50~ns) are omitted from the analysis.
The 50~ns window is determined from simulations, which showed that the width of air-shower signals is approximately 40-45~ns for the conditions of Tunka-Rex.  
\item \emph{Sliding noise window}.
As experience has shown the SNR estimation using a fixed noise window (slightly before the signal window in case of Tunka-Rex) is affected by occasional RFI in the noise window.
To improve this estimation we use a sliding window of 500~ns and define the noise level as the smallest RMS in the entire trace within the noise window.
Since this value is systematically smaller than the average noise level, the threshold SNR was increased from $10.0$ to the value of ${\mathrm{SNR_{th}}=16.0}$.
\item \emph{Neighborhood SNR}.
Since the approach described above is useful for treating RFI out of the signal window, the following cut is used in cases when RFI passes the signal window and/or overlaps with signal.
In this case the power of the signal peak is divided by the RMS of amplitudes surrounding the signal (i.e. $\pm100$~ns around the pulse).
Since we do not have the full description of all possible RFI, in this case, the conservative cut of ${\mathrm{SNR_{neighb.}}\ge10.0}$ is set for data analysis.
\item \emph{Manual RFI rejection}.
As mentioned before, after the Tunka-Rex upgrade, due to intensive developments in the valley many transient broadband peaks appeared in the Tunka-Rex traces.
The antenna stations spoiled with this noise were rejected manually.
In future we plan to train an artificial neural network for automatically tagging antennas contaminated by such RFI.
\end{itemize}

Since the true signal is heavily impacted by the noise at low SNRs it is necessary to take this into account.
First, SNR determines the uncertainty $\sigma(t',\mathrm{SNR})$ of the reconstructed signal.
Secondly, the noise contamination changes the total power of the signal:
on the one hand the total power is increased on average,
on the other hand it is decreased after applying a median filter, used to remove narrow-band RFI.
Therefore the signal is corrected using a function derived from simulations with measured background: ${u(t') \to u(t')(1 + f_c(t',\mathrm{SNR}))}$.
Details can be found in Refs.~\cite{Allan:1970xr,Schroeder2012S238}.
The novelty of our approach is the additional functional dependence on the time in the trace $t'$ relative to a peak of the signal ($t' = 0$ at the peak).
Thus the full pulse shape of the envelope is corrected for a bias due to noise (see Appendix~\ref{append:sigma_u} for details).

\section{Event reconstruction}
\begin{figure}
\includegraphics[width=0.5\linewidth]{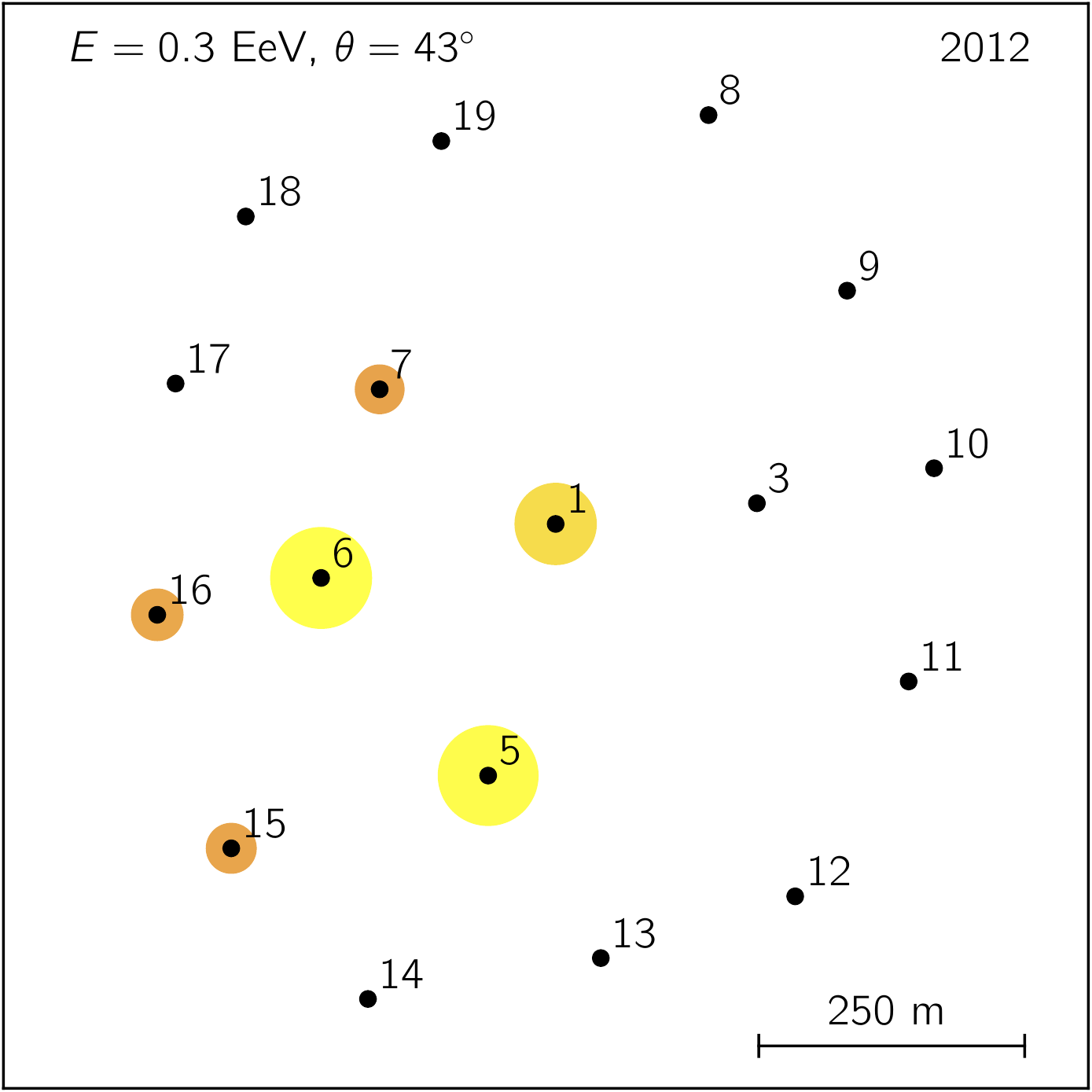}~\includegraphics[width=0.5\linewidth]{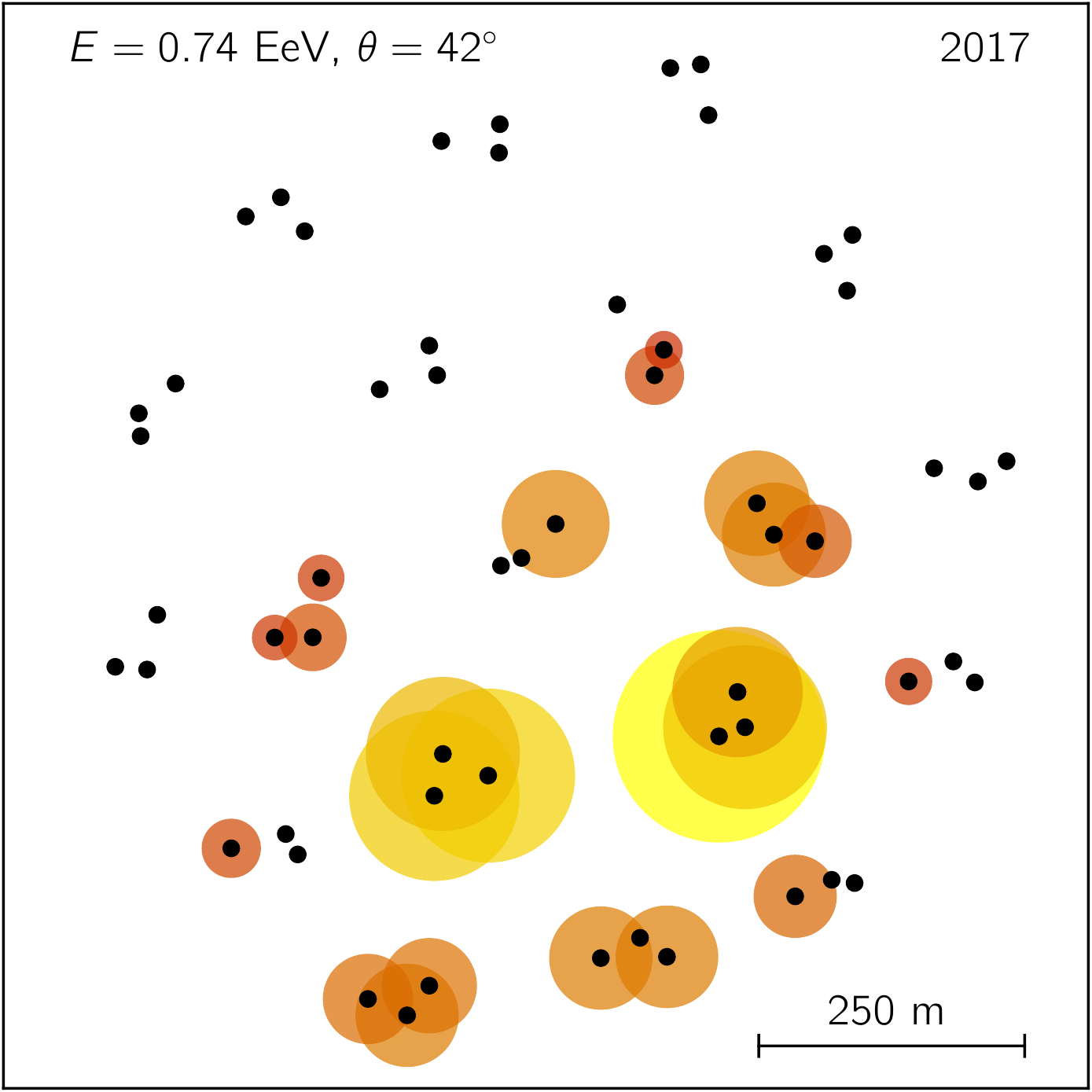}
\caption{Typical footprints of the detected air showers for the first detector configuration in 2012 (\emph{left}) and the latest in 2017 (\emph{right}).
During these years the density of Tunka-Rex was increased by three times.
The layout of Tunka-Rex is mostly determined by the infrastructure of the Tunka-133 and Tunka-Grande detectors.
Dots denote antenna stations, the size of the circles is proportional to amplitudes of radio signals (satellite antenna stations are not included for illustrative purposes).
The further reconstruction of the left event is depicted in Figs.~\ref{fig:signal_fit},~\ref{fig:xmax_chi2}.
}
\label{fig:event_footprint}
\end{figure}

The improved reconstruction of Tunka-Rex consists of the following basic steps:
\begin{enumerate}
\item Pre-reconstruction using the standard Tunka-Rex analysis pipeline with the improvements described above.
This reconstruction provides the shower axis and core position from Tunka-133 and the energy reconstructed by Tunka-Rex.
Let us note that since 2012 (generation \emph{1}) the density of the Tunka-Rex antenna array was increased threefold (starting from 2016, generation \emph{3}), yet the hardware and reconstruction pipeline are almost the same for all three generations (except for updates in calibration, data format, etc.).
The footprints of reconstructed example events from 2012 and 2017, respectively, are given in Fig.~\ref{fig:event_footprint}.
\item Creating a library with CoREAS simulations for each event obtained in the previous step with the goal to cover all possible depths of shower maxima possible for the particular event.
The reconstructed energy and geometry were used as input for the simulations with different primaries: hydrogen (proton), helium, nitrogen and iron nuclei~\footnote{The simulations and corresponding documentation will be published at \texttt{tunka.astroparticle.online} in the frame of Russian-German Data Life Cycle Initiative}.
We use CORSIKA v75600~\cite{HeckKnappCapdevielle1998} with QGSJet-II.04~\cite{Ostapchenko:2010vb}.
The selection of the hadronic model does not play a significant role in this analysis, for details see Appendix~\ref{append:hadronic_comp}.
\item Chi-square fit of the simulated envelopes against reconstructed ones.
The shower maximum and primary energy are reconstructed from the fits.
\end{enumerate}

In the present method the envelope of the electrical field $u(t)$ at the antenna station is used instead of the amplitude or power of the signal.
The measured amplitudes are prepared as follows:
each signal is bounded within the signal window $t_w<40$~ns (the width is flexible and optimized to have only one peak per envelope), and all of them are concatenated to a single time series $U(t)$ with length of ${N_b = N_a\cdot t_w / t_b}$, where $N_a$ is the number of antenna stations, $t_b$ is the size of bin:
\begin{equation}
U(t) = \bigoplus\limits_{i = 1}^{N_a} u_i(t')\Pi((t_i - t')/t_w)\,,
\end{equation}
where $t'$ is the index within trace $u_i$ at $i$-th antenna station, and $t_i$ is the position of the peak at this antenna station.
$\Pi(x)$ denotes the rectangular window function, where ${\Pi(-1/2<x<1/2) = 1}$.
The concatenated uncertainties $\sigma(t)$ of the measured signal (see Appendix~\ref{append:sigma_u}) and template $V(t)$ from the simulation are defined the same way.
It is worth to note, that for a more precise fit it is necessary to upsample the traces with resolutions ${t_b < 1}$~ns.
The technique of concatenation enables us to perform the following chi-square fit in a more elegant way.
Equivalently, the algorithm could iterate over all antenna stations with signal.

After concatenation of the traces the simulated electrical field $V(t)$ is fitted to a measured one $U(t)$ using one free parameter $A$, namely the normalization factor, using chi-square criteria:
\begin{equation}
\chi_{\mathrm{red.}}^2 = \sum\limits_{j = 1}^{N_b}\left(\frac{U_j - AV_j}{\sigma_j}\right)^2 \cdot \frac{f_{\mathrm{ups.}}}{N_\mathrm{bins}}\,.
\label{eq:signal_fit}
\end{equation}
The proper normalization of chi-square $\chi_{\mathrm{red.}}^2$ is defined by the upsampling factor $f_{\mathrm{ups.}}$ and our estimation is valid only for ${f_{\mathrm{ups.}} \gg 1}$ (${f_{\mathrm{ups.}} = 16}$ in our case).
The shower maximum $X_\mathrm{max}$ is defined as minimum of the parabola $\chi_{\mathrm{red.}}^2(X_\mathrm{max})$ with the confidence interval defined by the standard procedure ($\min(\chi_{\mathrm{red.}}^2) + 1$).

It is worth noticing that all chi-square distributions obtained in the present analysis are lying on the parabolic curves
while in Ref.~\cite{Buitink:2014eqa} only the points around minima are described by the parabola.
In the original paper~\cite{Buitink:2014eqa} this behavior is described as ``jitter'' introduced by shower-to-shower fluctuations which is also present in our analysis.
However our chi-square distributions still conserve the parabolic shape even far from the minimum which points to a different explanation of the ``jitter'' behavior obtained by LOFAR.

\begin{figure}
\includegraphics[width=1.0\linewidth]{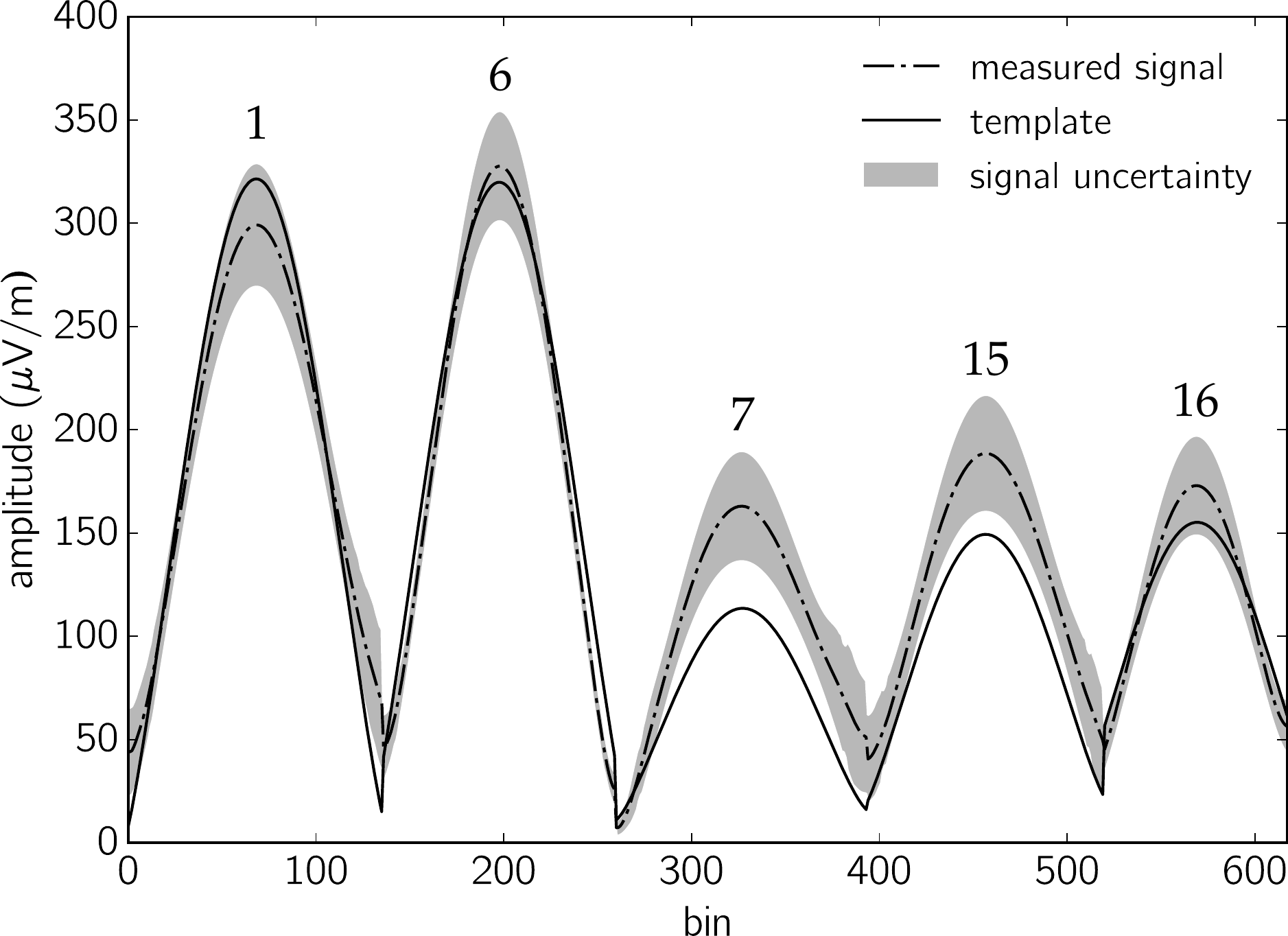}
\caption{Example of the concatenated trace described in Eq.~(\ref{eq:signal_fit}).
The normalization factor ${A = 0.975\pm0.006}$ is derived from the fit of the template $V(t)$ (solid line) to the measured signal $U(t)$ (dashed line) with uncertainties $\sigma(t)$ (gray area).
The template is scaled by the factor $A$ for illustration purposes.
The fit is performed over ${N_b = 618}$ bins for ${N_a = 5}$ antenna stations, i.e. each pulse is described by about 128 bins with binning of ${t_b = 0.3125}$~ns.
The numbers on top of peaks correspond to antenna stations in Fig.~\ref{fig:event_footprint} (left).
The antenna station 5 is omitted in this analysis due to technical reasons (new coordinates after pole replacement were not included in analysis).}
\label{fig:signal_fit}
\end{figure}

\begin{figure}
\includegraphics[width=1.0\linewidth]{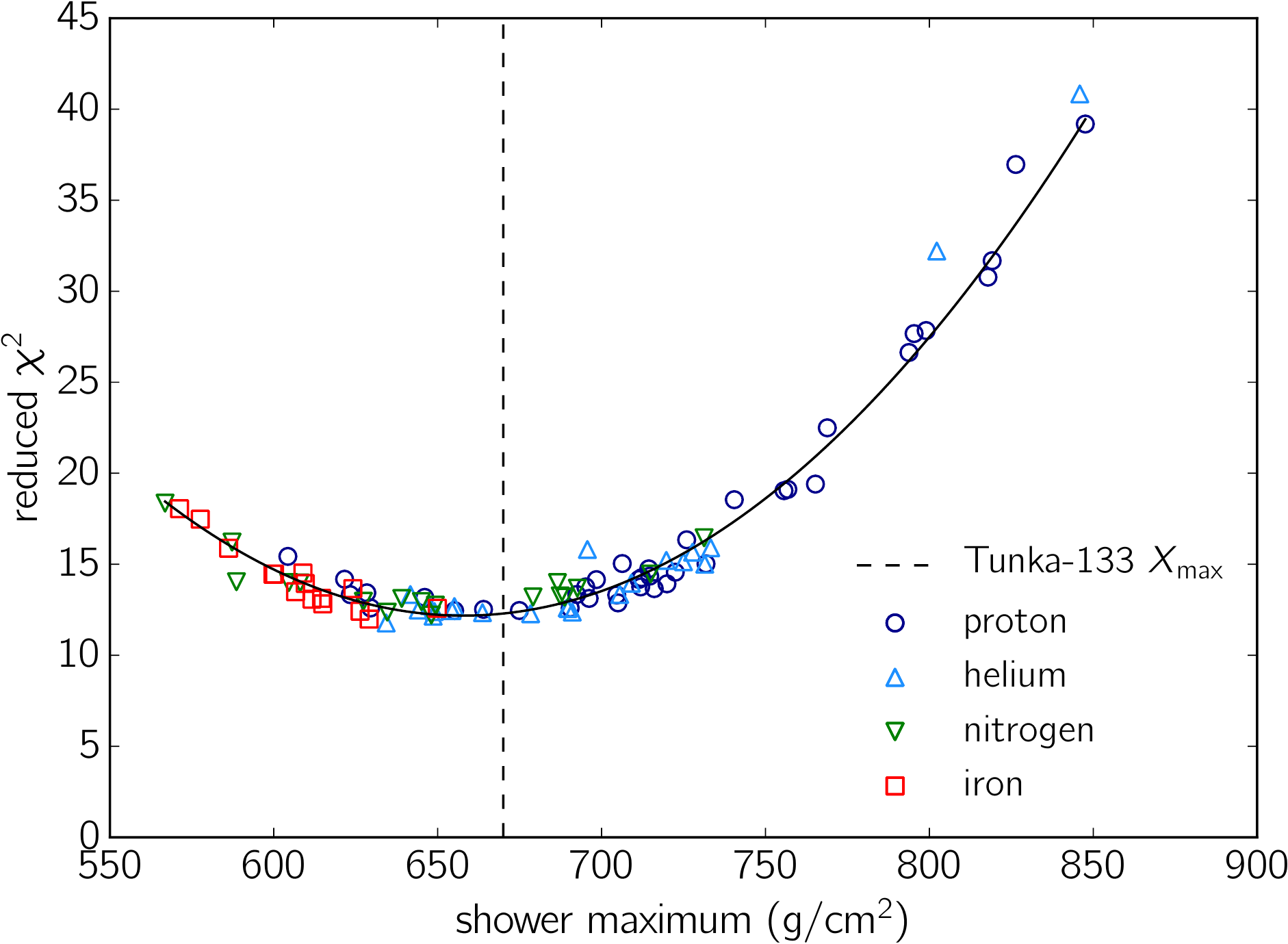}
\caption{The example $\chi_{\mathrm{red.}}^2(X_\mathrm{max})$ distribution for the shower maximum reconstruction.
Tunka-Rex reconstructed ${X_\mathrm{max} = 658\pm36}$~g/cm\textsuperscript{2} (minimum of parabola), Tunka-133 ${X_\mathrm{max} = 670\pm25}$~g/cm\textsuperscript{2} (shown by dashed line).
The reduced $\chi^2$ does not equal unity since the uncertainties of the hardware model and simulations are not included in the fit.
However, estimating the width of the reduced $\chi^2$ of ${\sigma_{\chi^2_{\mathrm{red.}}}=\sqrt{2N_\mathrm{bins}/f_{\mathrm{ups.}}}\approx9}$, one can see that the deviation from unity is of only about one sigma.
As expected from shower universality it is not possible to resolve the single primary particle using only measurement of electromagnetic components (all four primaries have the same $\chi^2$ values around the minimum of the parabola).}
\label{fig:xmax_chi2}
\end{figure}

An example of a single fit is given in Fig.~\ref{fig:signal_fit}, and the $\chi_{\mathrm{red.}}^2(X_\mathrm{max})$ distribution for this event is shown in Fig.~\ref{fig:xmax_chi2}.

\section{Event selection and efficiency cuts}
\label{sec:efficiency}
To reconstruct the flux and mean depth of shower maximum of primary cosmic rays one has to take into account the efficiency of detection.
There are different ways to estimate this, for example based on simulations as used by LOFAR~\cite{Buitink:2016nkf}.
We developed our own, simple model based on Monte-Carlo simulations of air-shower footprints, a detailed description of which is given in Ref.~\cite{Fedorov_TunkaRex_ICRC2017}.
The idea behind this model is to estimate the probability of detection of an air shower with particular energy and direction (this way we do not neglect the azimuth of arrival since it impacts the geomagnetic and charge-excess effects) with a random shower core location inside the effective detection area.
The efficiency is calculated as the fraction of showers detected inside the detector area, a circle with radius $R_\mathrm{detector}$ around the center of the array.
The detection criterium selected for this work is to have a footprint covering at least four clusters of Tunka-Rex.
This criterium guarantees the detection of the air showers with different depths, i.e. minimizes the bias due to nondetection of deep proton showers, for example.
We use the following cut: 90\% efficiency of detection of air-showers which produce footprints with size of $R_\mathrm{footprint}\approx300$~m (depending on shower inclination; footprints have the shape either of a circle or of an ellipse.
In case of an ellipse, $R_\mathrm{footprint}$ corresponds to the minor axis) within the area of ${R_\mathrm{detector} = 450}$~m around the center of the antenna array.

To validate this model we performed the following: for each generation of the Tunka-Rex array we predict the expected number of detected events with 90\% efficiency using the information (energy, arrival direction and core position) from Tunka-133 (assuming that Tunka-133 features full efficiency, which is valid at these energies) and compared these numbers with events detected and reconstructed by Tunka-Rex.
The comparison is given in Table~\ref{tab:eff_comparison}.
From this table, one can see that the number of detected events are in agreement with the number of predicted ones within the pre-defined efficiency of detection.
Since the number of missed events is small all of them were revised manually and it was found that the core positions of these events are located within the 10\% of ``blind'' parts of the detector, as predicted by the efficiency model.
This is especially true for the generation \emph{1a} of the array, where the structure is more irregular due to absence of antennas in two clusters (see Fig.~\ref{fig:event_footprint}, left).
The efficiency of 94\% determined by the cross-check is consistent with the cut condition of $>90$\%.

\begin{table}[h!]
\caption{Comparison between the expected number of events as predicted by the model of efficiency with the number of events detected and reconstructed by Tunka-Rex.}
\label{tab:eff_comparison}
\begin{ruledtabular}
\begin{tabular}{llllll}
Gen.   & Years     & Number of        & Expected        & Detected        & Efficiency \\
       &           & antennas         & events          & events          &            \\
\hline
1a     & 2012/13   & 18               & 23              & 20              & $0.85_{-0.09}^{+0.05}$ \\
1b     & 2013/14   & 25               & 28              & 27              & $0.96_{-0.05}^{+0.02}$ \\
2      & 2015/16   & 44               & 14              & 14              & $1.00_{-0.07}^{+0.00}$ \\
3      & 2016/17   & 63               & 17              & 16              & $0.94_{-0.08}^{+0.04}$ \\
\hline
       &           & Total            & 82              & 77              & $0.94_{-0.03}^{+0.02}$ \\
\end{tabular}
\end{ruledtabular}
\end{table}

\section{Quality of reconstruction}
To derive the quality and precision of the Tunka-Rex reconstruction we cross-check it with the Tunka-133 reconstruction.
Estimating the Tunka-133 resolution for the energy and shower maximum as approximately 10\% and 25~g/cm\textsuperscript{2}, respectively, one can estimate the resolution of Tunka-Rex by analyzing the deviation between the values estimated by the two detectors.
For the events presented in Table~\ref{tab:eff_comparison}, the average energy resolution of Tunka-Rex is about 10\% (here it is worth noting that since both detectors reconstruct the energy via the electromagnetic component, an additional uncertainty caused by mass composition is compensated).
As was shown before~\cite{Kostunin:2016xbk} the sensitivity to the shower maximum increases with the size of the radio footprint of the air-shower.
With the increased number of antenna stations and sufficient number of events we studied the dependency of the $X_\mathrm{max}$ resolution on the primary energy (which impacts the number of antennas per events and size of the air-shower footprint), and found that we can achieve a resolution similar to the resolution of Tunka-133 (and other optical detectors), for energies above $10^{17.7}$~eV.
At these energies the air-shower footprint of the typical event has a radius of about 450~m in our zenith angle range $<50^\circ$.
Referring to LOFAR results, one can state that the number of antenna stations is the more important factor than the size of the footprint.
In the present work it is hard to judge whether the increased density of Tunka-Rex antenna stations significantly improves the reconstruction, since the array with tripled density has not acquired sufficient data yet and only Tunka-133 triggered events are fully analyzed.
In Table~\ref{tab:resolution} one can see the comparison of shower maximum resolutions of Tunka-Rex achieved at different energy cuts.
These values are in agreement with a mean uncertainty of $31$~g/cm\textsuperscript{2} given by the chi-square fit.

It is important to note that after the commissioning of new gamma instruments in the Tunka Valley,
the maintenance and upgrades of Tunka-133 were less regular,
and the present data analysis finishes in 2016.
The analysis of the latest data is not included in the Tunka-133 spectra and mass composition, i.e. the energy and $X_\mathrm{max}$ values provided for the Tunka-Rex cross-check are preliminary and might change in the future.
Although the reconstruction of Tunka-Rex is valid for 2016/17 and later, the cross-calibration of this season against Tunka-133 is not performed in this paper.

\begin{table}[h!]
\caption{Resolution of Tunka-Rex on the shower maximum as derived from a cross-check with Tunka-133 as a function of primary energy.
The resolution is derived from the difference between relative deviations (see Fig.~\ref{fig:resolutions}) and Tunka-133 resolution itself using the standard formula for propagation of uncertainty.
The mean uncertainty given by chi-square fit is about $31\pm2$~g/cm\textsuperscript{2} and constant with energy.}
\label{tab:resolution}
\begin{ruledtabular}
\begin{tabular}{llll}
$E_{\mathrm{pr}}$ (eV)   & $R_{\mathrm{footprint}}$ (m)  & $\langle N_{ant}\rangle$/event      & $\sigma{X_\mathrm{max}}$ (g/cm\textsuperscript{2}) \\
$<10^{17.5}$             & $>240$        & 5    & $\ge 30$ \\
$10^{17.5}$ -- $10^{17.7}$ & $>320$        & 6    & $\approx 30$ \\
$>10^{17.7}$             & $>430$        & 7    & $\le 25$ \\
\end{tabular}
\end{ruledtabular}
\end{table}

\section{Discussion and conclusion}
\begin{figure}[h!]
\includegraphics[width=1.0\linewidth]{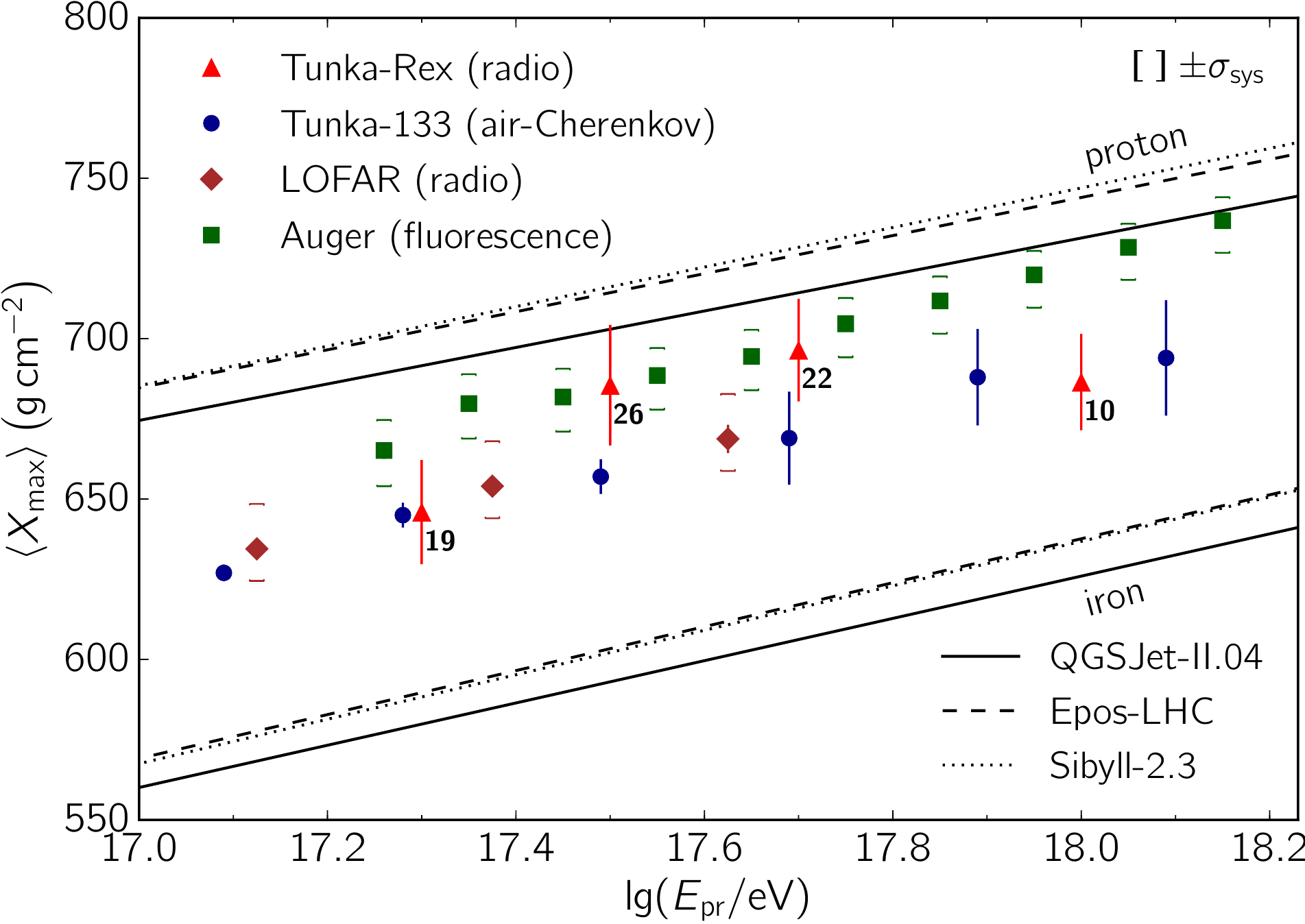}
\caption{Mean atmospheric depth of shower maximum $X_\mathrm{max}$ as a function of energy reconstructed by Tunka-Rex.
Error-bars indicate statistical uncertainty of the mean (using standard definition).
The numerical values of the points and distribution inside bins are given in Appendix~\ref{append:distr}.
The values of other experiments are taken from Refs.~\cite{Prosin:2016jev,Buitink:2016nkf,Bellido:2017cgf}, the model curves are from Refs.~\cite{Ostapchenko:2010vb,Pierog:2006qv,Riehn:2015oba}.
}
\label{fig:xmax_spectrum}
\end{figure}

As another check of the air-shower reconstruction in combination with the efficiency model we produced the distribution of the mean depth of the reconstructed shower maxima as a function of the primary energy.
The results are presented in Fig.~\ref{fig:xmax_spectrum}.
One can see that the Tunka-Rex results are in agreement with other experiments.
The points presented in this plot are obtained with three different techniques: air-Cherenkov, fluorescence and radio.
The good agreement between the three techniques shows the progress in the understanding of air-shower phenomena and systematics of experiments exploiting these techniques.

We made a more detailed comparison with the Tunka-133 reconstruction on a large data set from 2012-2015 and found that Tunka-Rex shows mean shifts against Tunka-133 of $-2\pm1$\% and $12\pm4$~g/cm\textsuperscript{2} for energy and shower maximum (depending on energy, see Fig.~\ref{fig:resolutions}).
This can be explained by unaccounted systematics due to degradation of the Tunka-133 optical system and different methods of reconstruction (earlier Tunka-Rex used an LDF method~\cite{Kostunin:2015taa,Bezyazeekov:2015ica} which is very similar to the one used by Tunka-133).

An unaccounted systematic uncertainty is given by the variation of the refractive index of air.
It was shown that the atmosphere can introduce a bias up to 11~g/cm\textsuperscript{2} for MHz radio emission~\cite{Corstanje:2017djm}.
Probably this effect is mostly compensated for the relative comparison of Tunka-Rex and Tunka-133, but can impact the absolute value of the shower maximum.
Since the software used in Ref.~\cite{Corstanje:2017djm} was implemented in the CORSIKA/CoREAS packages after we performed our analysis we provide only a rough estimation of systematic uncertainty introduced by the refractivity of the atmosphere in the present paper.
Using GDAS (Global Data Assimilation System)~\footnote{National Oceanic and Atmospheric Administration, https://www.ncdc.noaa.gov/data-access/model-data/model-datasets/global-data-assimilation-system-gdas} profiles for each detected event and formulas from Ref.~\cite{Corstanje:2017djm} the expected uncertainty is about 3~g/cm\textsuperscript{2} (event-to-event fluctuation of the refractivity is about 2\%) with a possible shift up to 5~g/cm\textsuperscript{2} towards deeper shower maxima (on average GDAS refractivity is 5\% higher than CORSIKA standard one).
These small-scale deviations are determined by the operation conditions of the Tunka-133 detector: dry and cold atmosphere with almost constant temperature.

The main results of this work can be summarized as follows:
\begin{itemize}
\item The simple efficiency model developed for Tunka-Rex shows good agreement with measurements.
\item Using a simulation-based approach and pulse shape information the resolution of the Tunka-Rex reconstruction was significantly increased for both the energy and shower maximum.
\item With the improved analysis method the radio measurements reach a resolution comparable to optical detectors; where radio has the intrinsic advantage of a higher duty cycle.
\end{itemize}

There are still a number of improvements which can be implemented.
First,  reducing the systematic uncertainty by taking realistic atmospheric conditions into account.
Second, improving the understanding of the phase response in order to use the shape of the signal itself and not just the envelope.
There is an order of magnitude larger amount of data acquired jointly with Tunka-Grande during daytime measurements.
On the one hand, analysis of these data requires a slightly different approach, on the other the data also contain information regarding the muonic component of air-showers which can dramatically improve the reconstruction of the primary mass.
This amount of data would be sufficient for a precise reconstruction of the flux of ultra-high energy cosmic rays, i.e. the energy spectrum.


\appendix
\section{Signal uncertainty due to noise and its adjustment}
\label{append:sigma_u}

\begin{figure}
\includegraphics[width=0.49\linewidth]{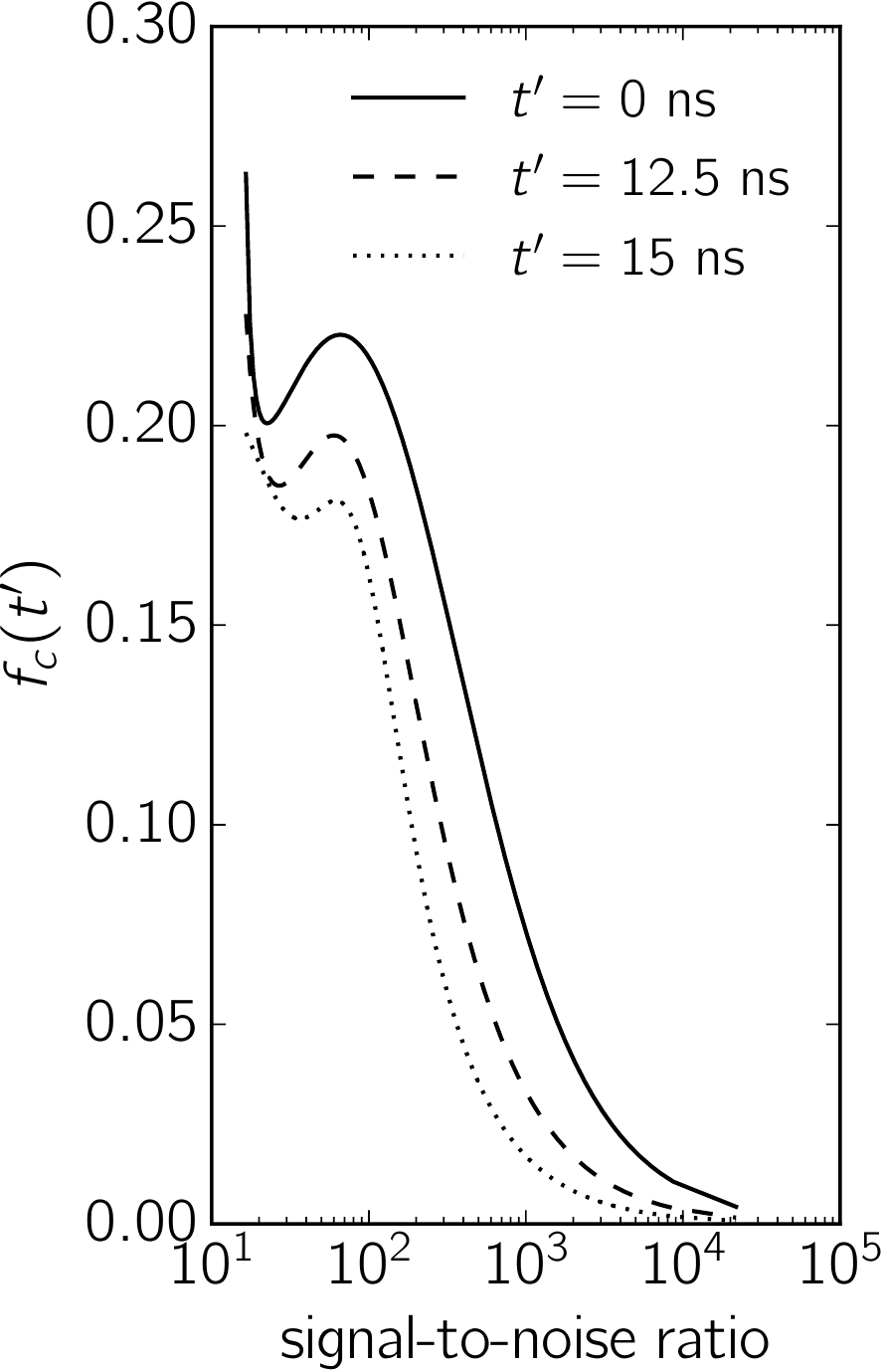}
\includegraphics[width=0.49\linewidth]{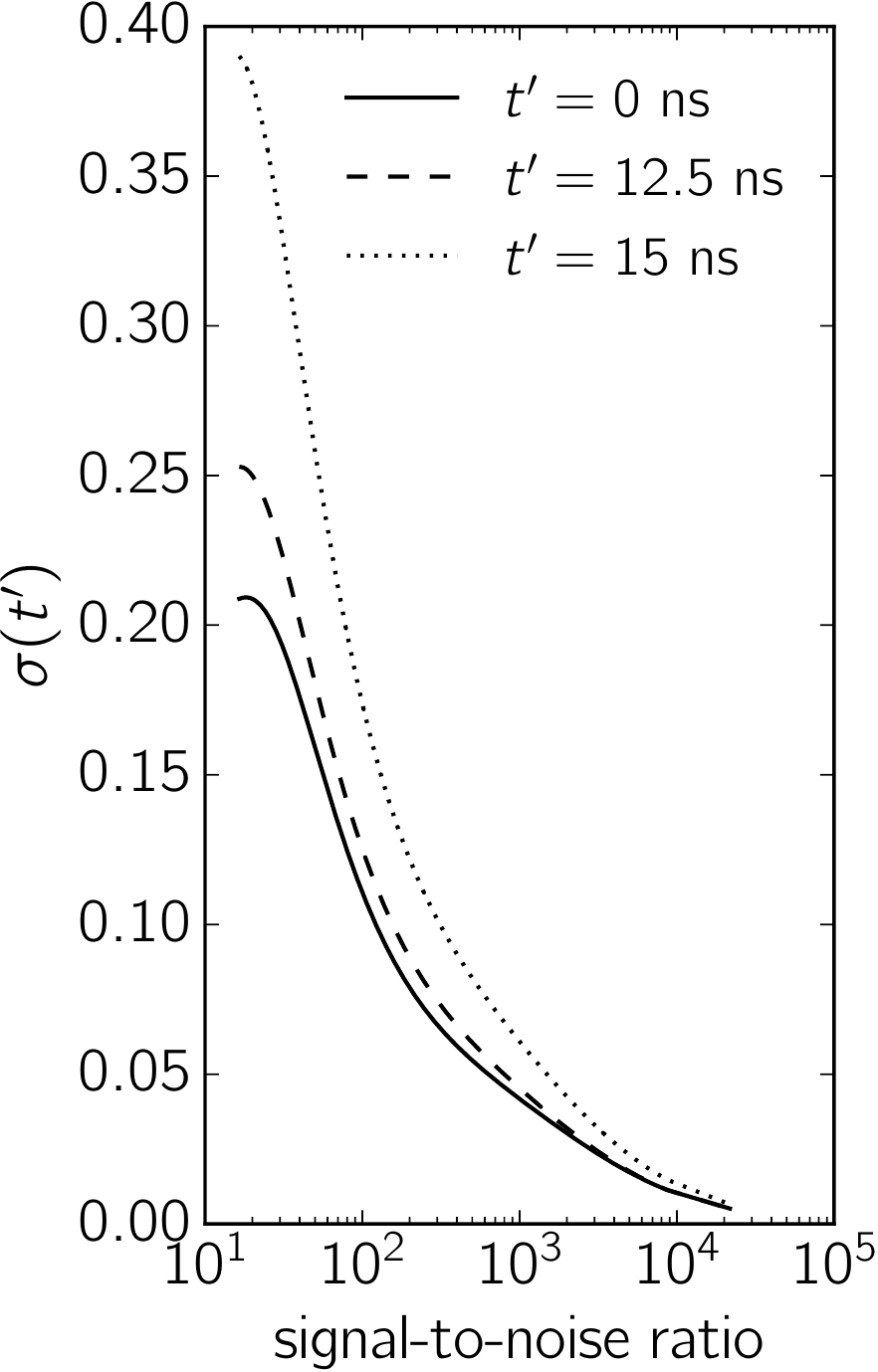}
\caption{The family of curves depicting the behavior of the functions $f_c(t',\mathrm{SNR})$ and $\sigma(t',\mathrm{SNR})$.}
\label{fig:signal_corr}
\end{figure}

As one can see from Fig.~\ref{fig:signals} the signal envelopes can be considered symmetric, i.e. we assume ${u(-t') = u(t')}$.
For simplicity both the signal uncertainty $\sigma(t',\mathrm{SNR})$ and the correction term $f_c(t',\mathrm{SNR})$ are parametrized as a sum of two Lorentzian-like functions:
\begin{eqnarray}
&&\sigma,f_c(t',\mathrm{SNR}) = L_1^{\sigma,f_c}(t',\mathrm{SNR}) + L_2^{\sigma,f_c}(t',\mathrm{SNR})\,,\\
&&L_{1,2}^{\sigma,f_c}(t',\mathrm{SNR}) = \frac{a_{1,2}^{\sigma,f_c}(t')\cdot\mathrm{SNR}}{\left(\mathrm{SNR}-b^{\sigma,f_c}_{1,2}(t')\right)^2 + c_{1,2}^{\sigma,f_c}(t')}\,.
\end{eqnarray}
The parameters $a_{1,2}^{\sigma,f_c}$, $b_{1,2}^{\sigma,f_c}$, $c_{1,2}^{\sigma,f_c}$ are obtained from the simulations after adding the measured noise samples from the Tunka-Rex noise library.
The behavior of the $\sigma(t',\mathrm{SNR})$ and $f_c(t',\mathrm{SNR})$ functions is depicted in Fig~\ref{fig:signal_corr}.

\section{Comparison of handronic models}
\label{append:hadronic_comp}

\begin{figure}[h!]
\includegraphics[width=1.0\linewidth]{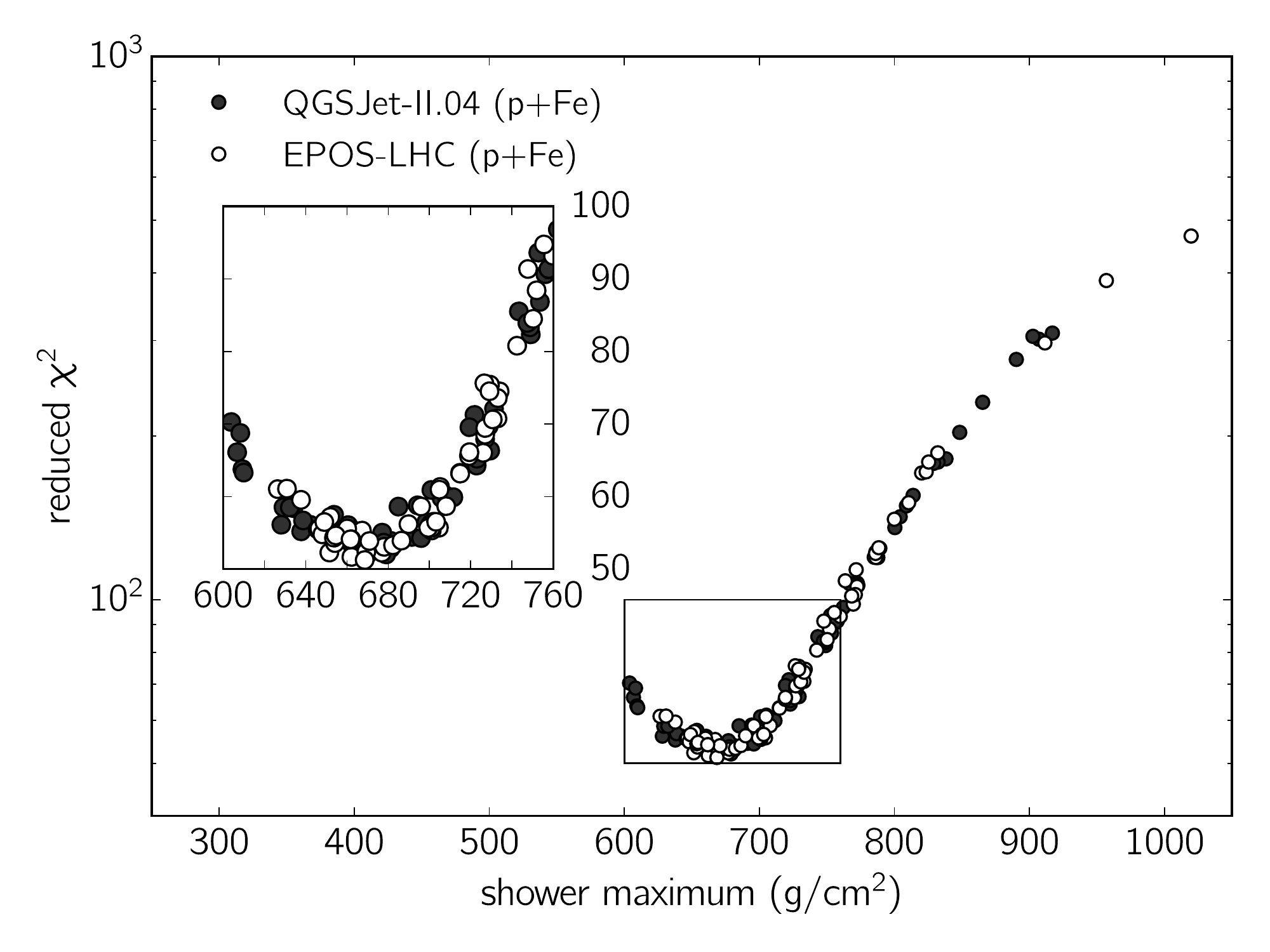}
\caption{Chi-square distribution produced by QGSJet-II.04 and EPOS-LHC models for the same event.
The depths of the shower maxima reconstructed from these distributions are $664\pm15$ and $662\pm15$ g/cm\textsuperscript{2}, respectively.
The shape of both parabolas are the same, but the distributions and densities of the points are different (see Fig.~\ref{fig:xmax_models_hist}).}
\label{fig:xmax_vs_chi2}
\end{figure}

\begin{figure}[h!]
\includegraphics[width=0.7\linewidth]{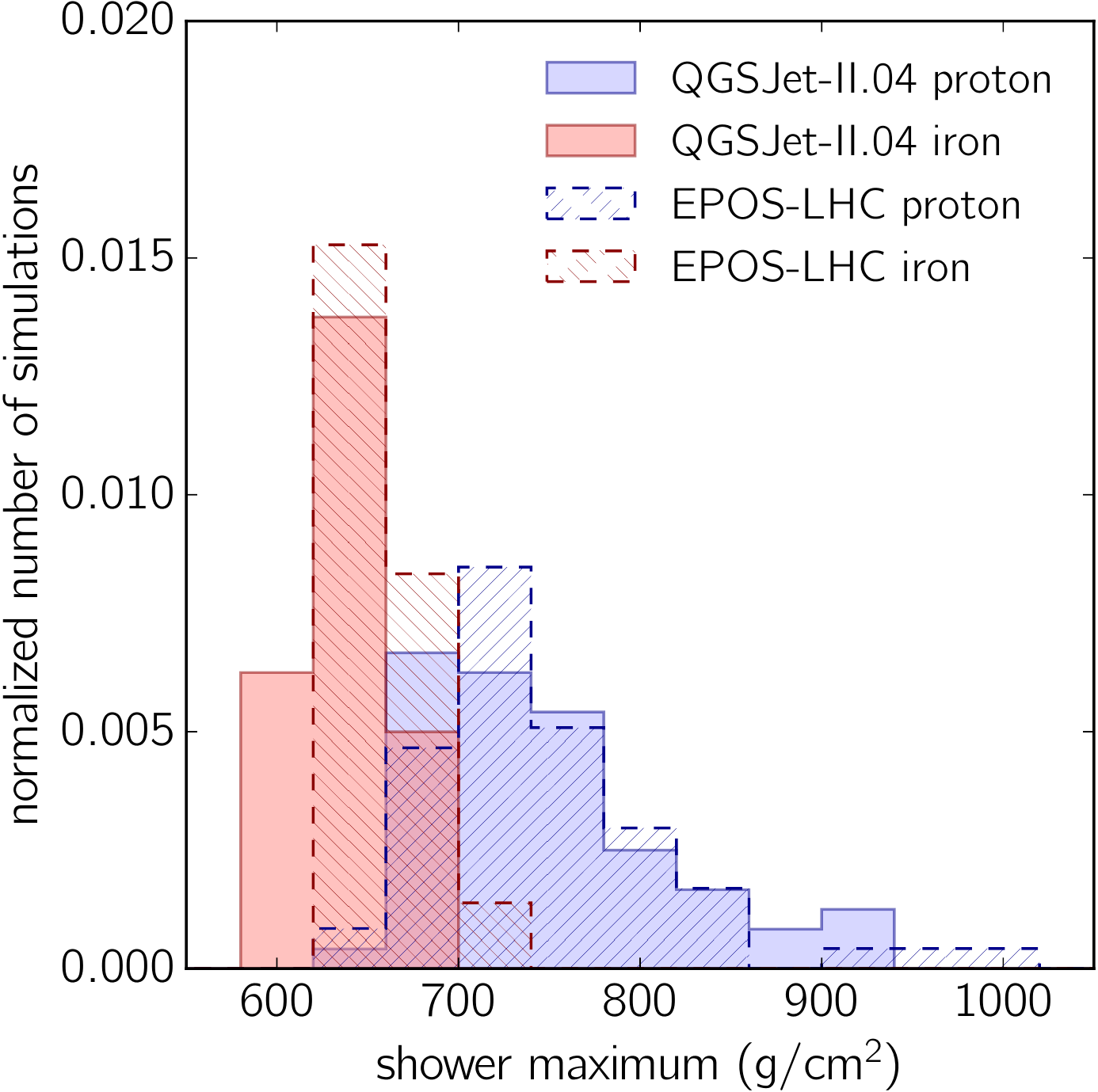}
\caption{The distribution of shower maxima generated by QGSJet-II.04 and EPOS-LHC models for the event shown in Fig.~\ref{fig:xmax_vs_chi2}.
The distributions systematically shifted against each other, however it has no impact on shower maximum reconstruction.}
\label{fig:xmax_models_hist}
\end{figure}

\begin{figure}[h!]
\includegraphics[width=0.9\linewidth]{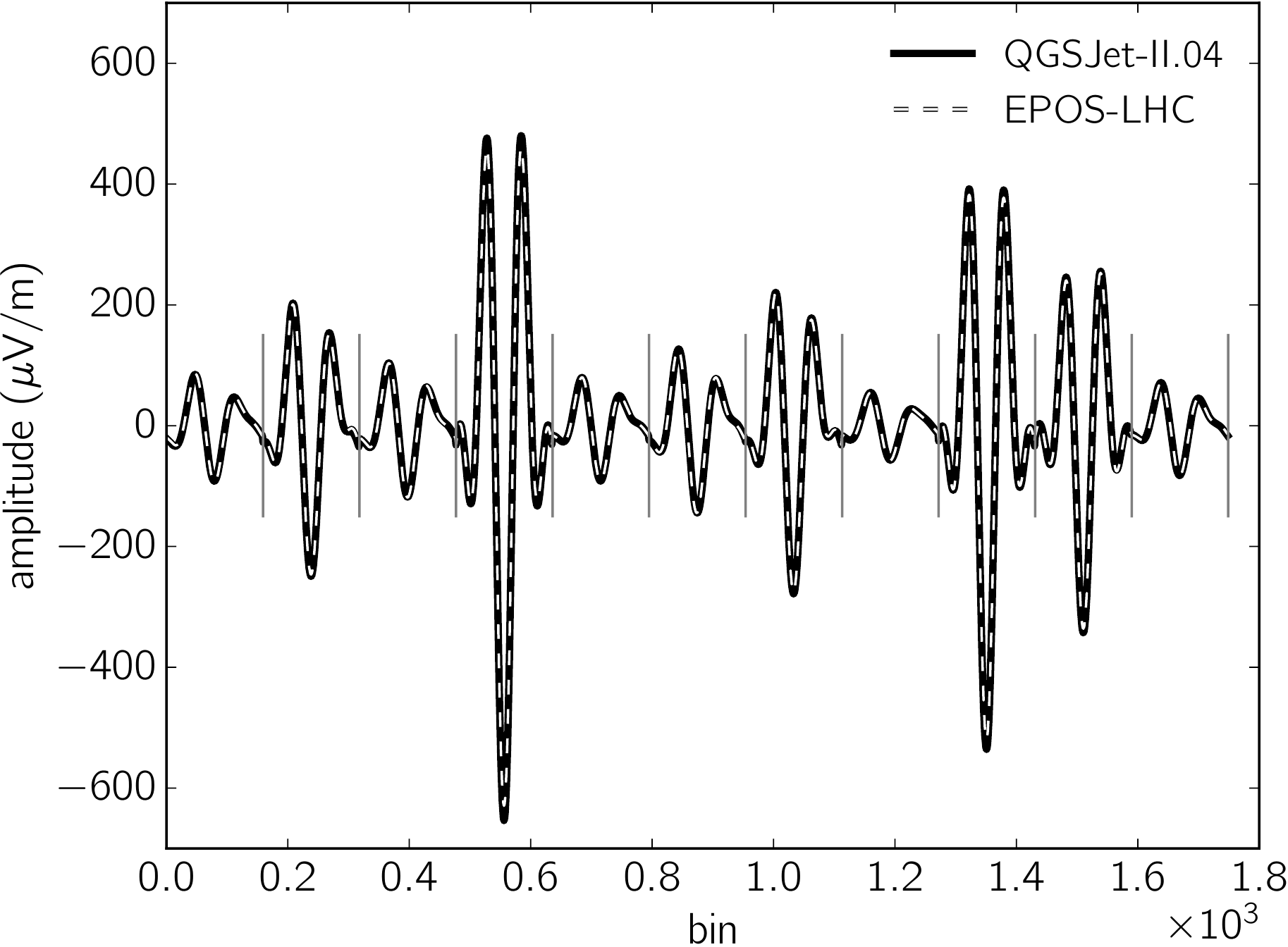}
\caption{Comparison of electric fields generated by QGSJet-II.04 and EPOS-LHC models for a similar event (identical primary particle and energy, shower maxima differ less than 1 g/cm\textsuperscript{2}).
Vertical lines separate signals at different observation points.
One can see no significant difference in pulse shapes and amplitudes which explains the result shown in Fig.~\ref{fig:xmax_vs_chi2}.}
\label{fig:traces_hadr}
\end{figure}

The selection of the hadronic model does not play a significant role in our analysis.
The physical explanation of this is the manner of generation of the electromagnetic component of the air shower.
The primary contribution comes from the production and immediate decay (due to short mean life) of neutral pions, which is calculated similarly by all models (the difference in pion production important for the muon number does not significantly impact radio emission, since fraction of muons in the total number of charged particles less than percent).
To explicitly show this in the frame of our reconstruction we give an example of the chi-square distribution for an example event reconstructed by QGSJet-II.04~\cite{Ostapchenko:2010vb} and EPOS-LHC models~\cite{Pierog:2006qv}.
One can see in Figs.~\ref{fig:xmax_vs_chi2}~and~\ref{fig:xmax_models_hist} that although both models provide different distributions of shower maxima the chi-square distributions fitted to them are lying in the same parabola.
In Fig.~\ref{fig:traces_hadr} one can see that electric fields produced by QGSJet-II.04 and EPOS-LHC for similar events (same primary particle, energy and similar shower maxima) are in very good agreement.

As has already been shown in Ref.~\cite{Buitink:2014eqa}, the difference in reconstruction of shower maxima obtained by different hadronic models is negligible.
Here we add that this difference might appear due to statistical fluctuations which are reflected in the chi-square fit, and might vanish with a sufficient number of simulations.

\section{Cross-check with Tunka-133}
\label{append:cross_check}

\begin{figure}
\includegraphics[width=0.49\linewidth]{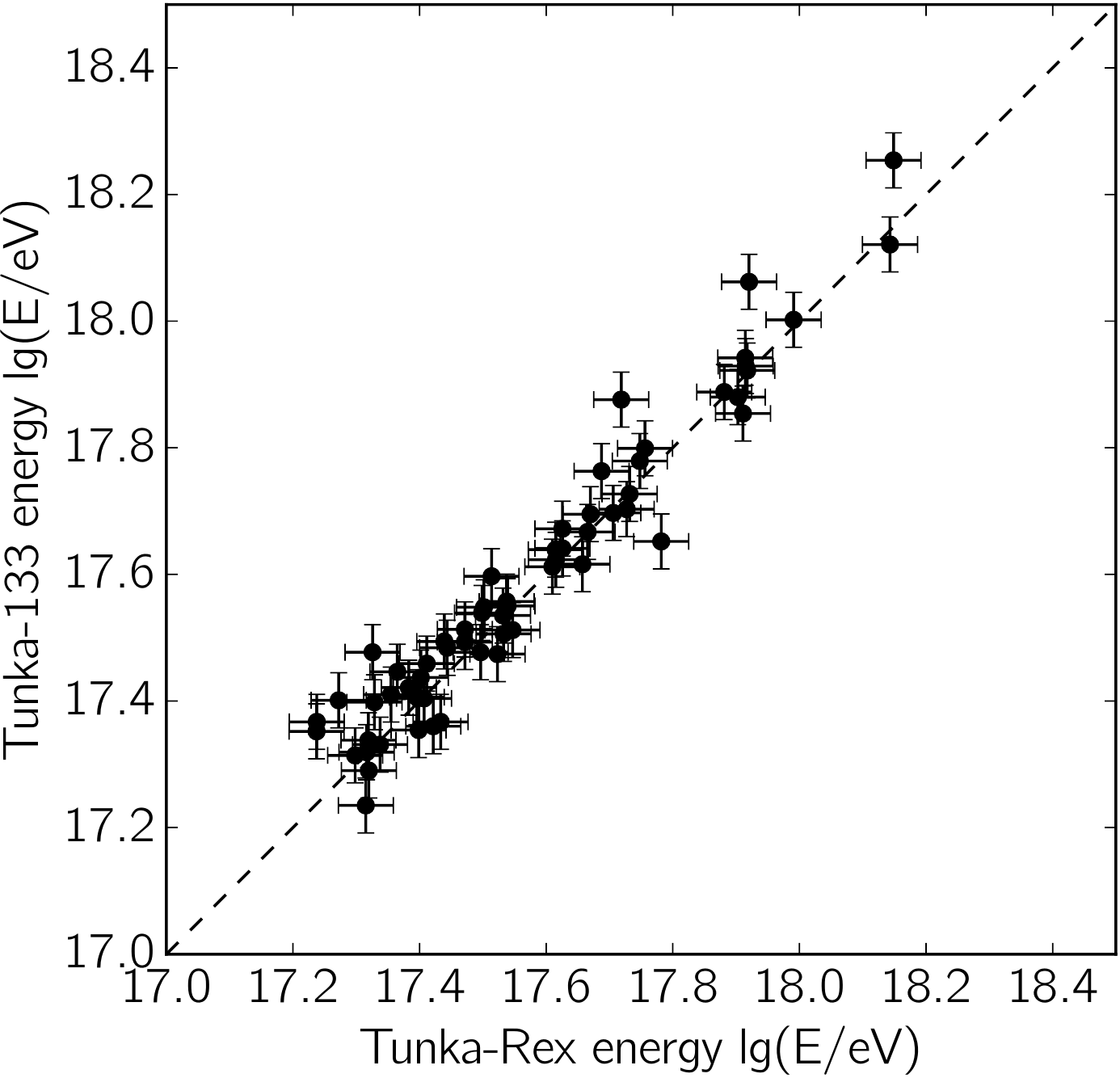}
\includegraphics[width=0.49\linewidth]{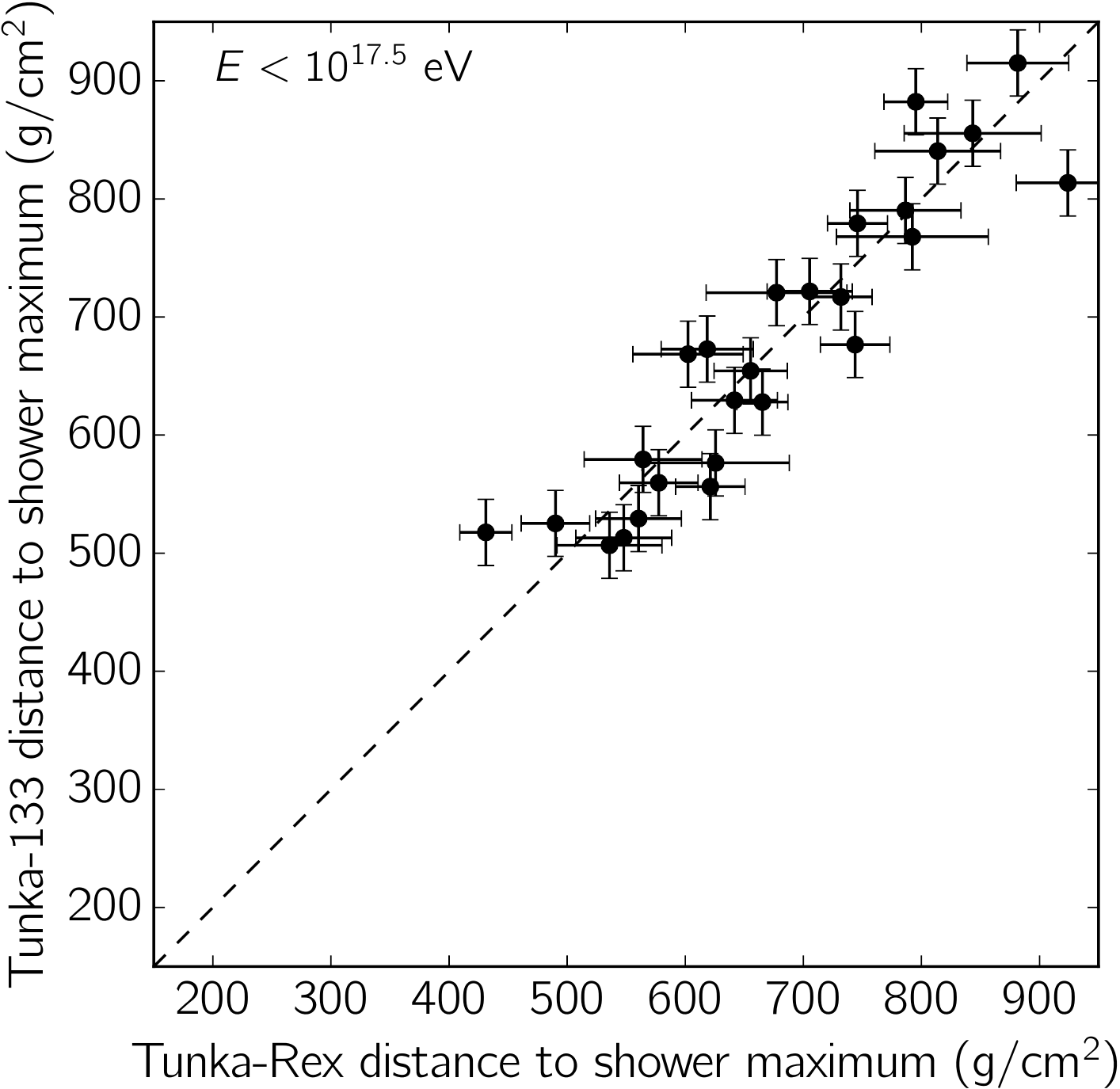}\\
~\\
\includegraphics[width=0.49\linewidth]{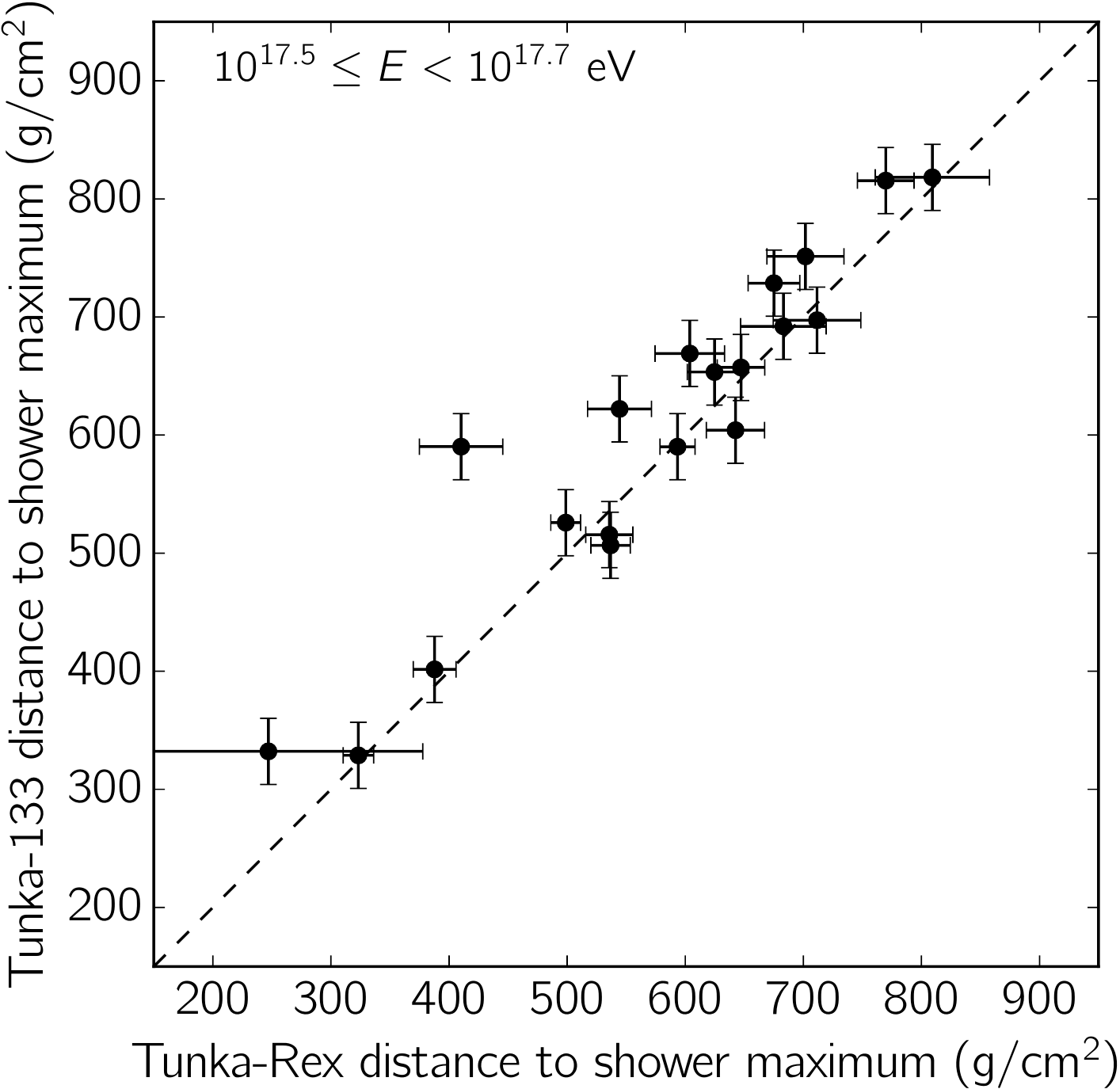}
\includegraphics[width=0.49\linewidth]{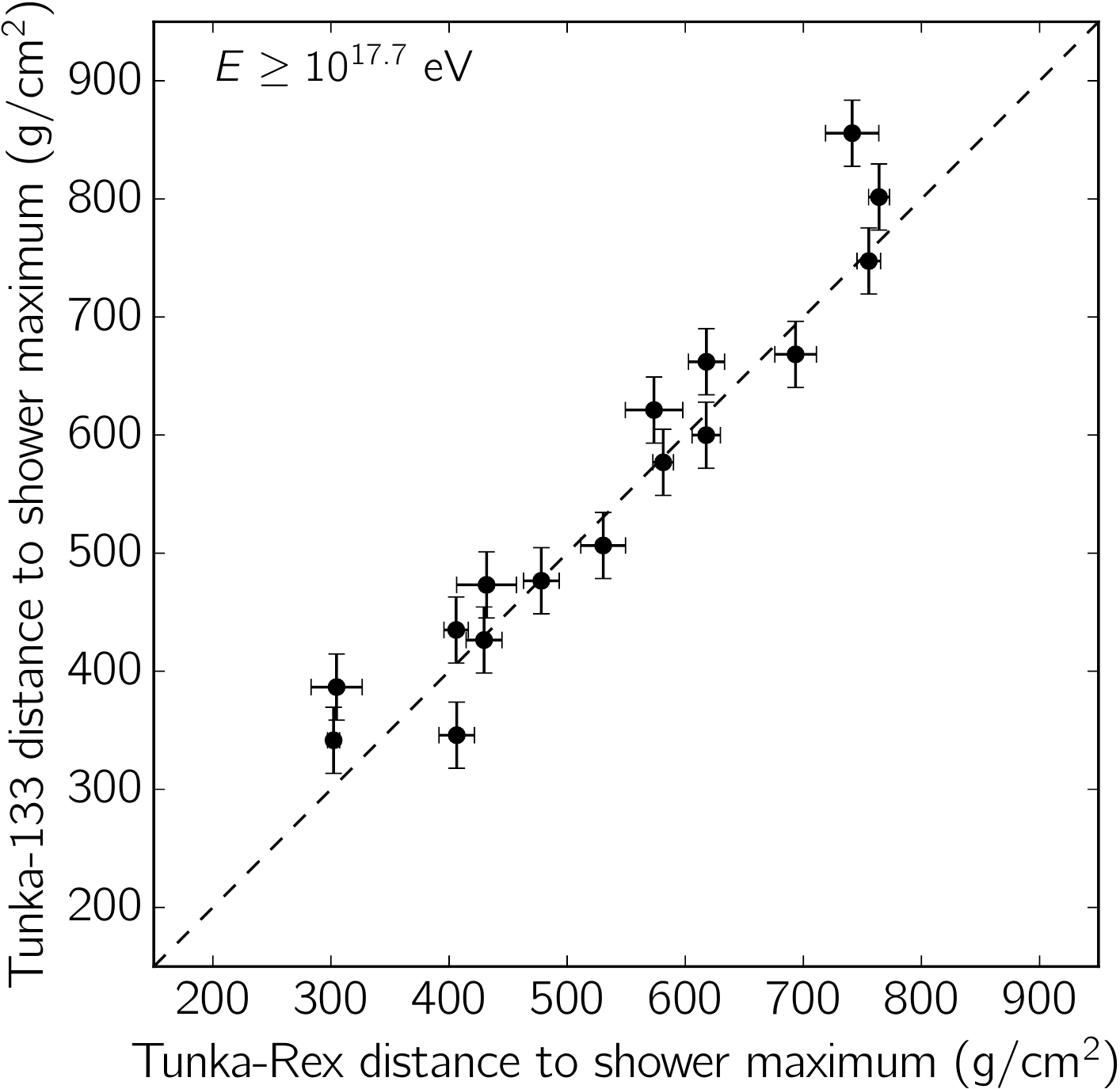}
\caption{Correlation of the energy (top left) reconstructed by Tunka-Rex and Tunka-133 and the reconstruction of shower maximum for different energy cuts.}
\label{fig:correlations}
\end{figure}

\begin{figure}
$E<10^{17.5}$~eV\\
\includegraphics[width=0.49\linewidth]{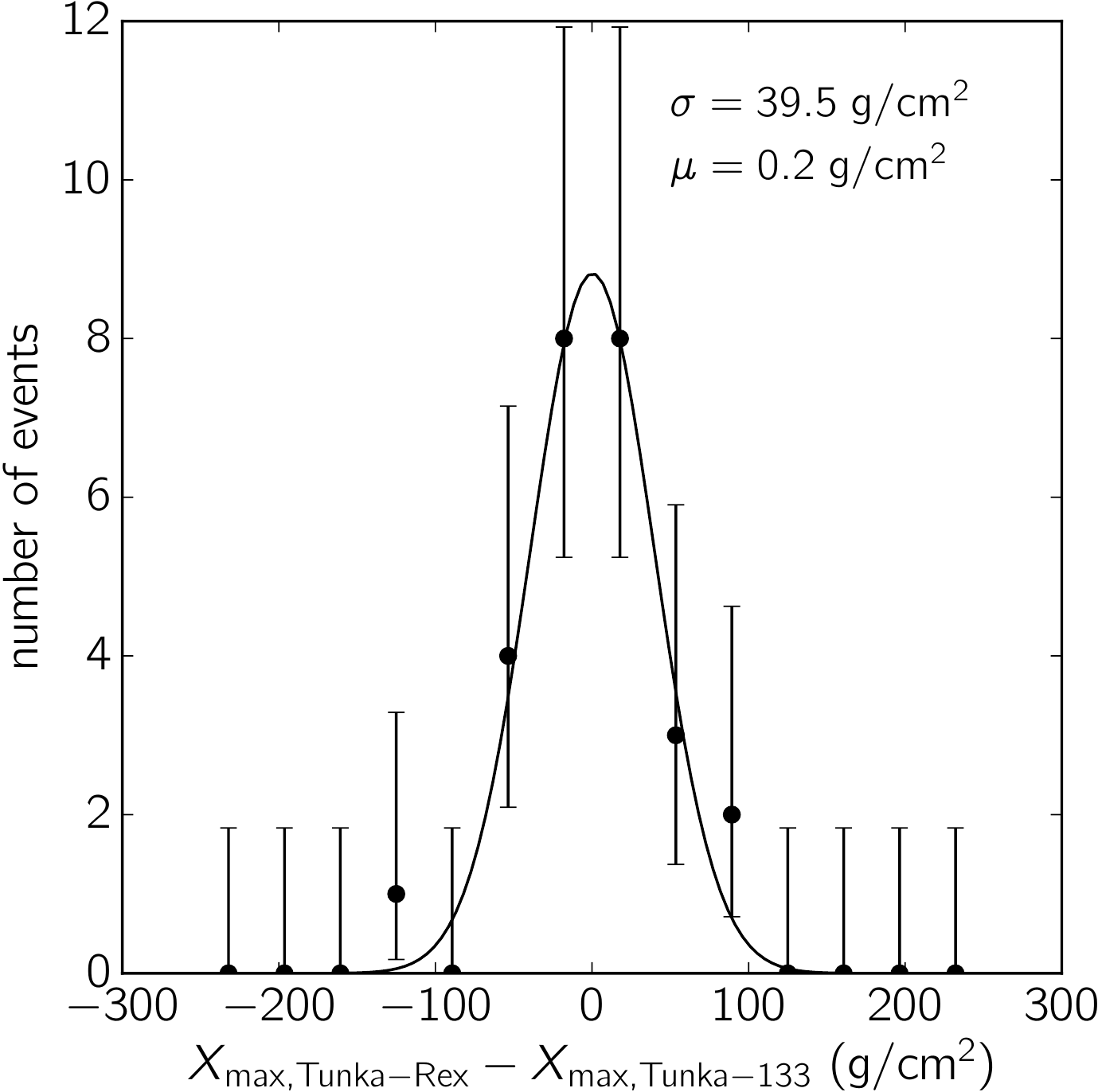}
\includegraphics[width=0.49\linewidth]{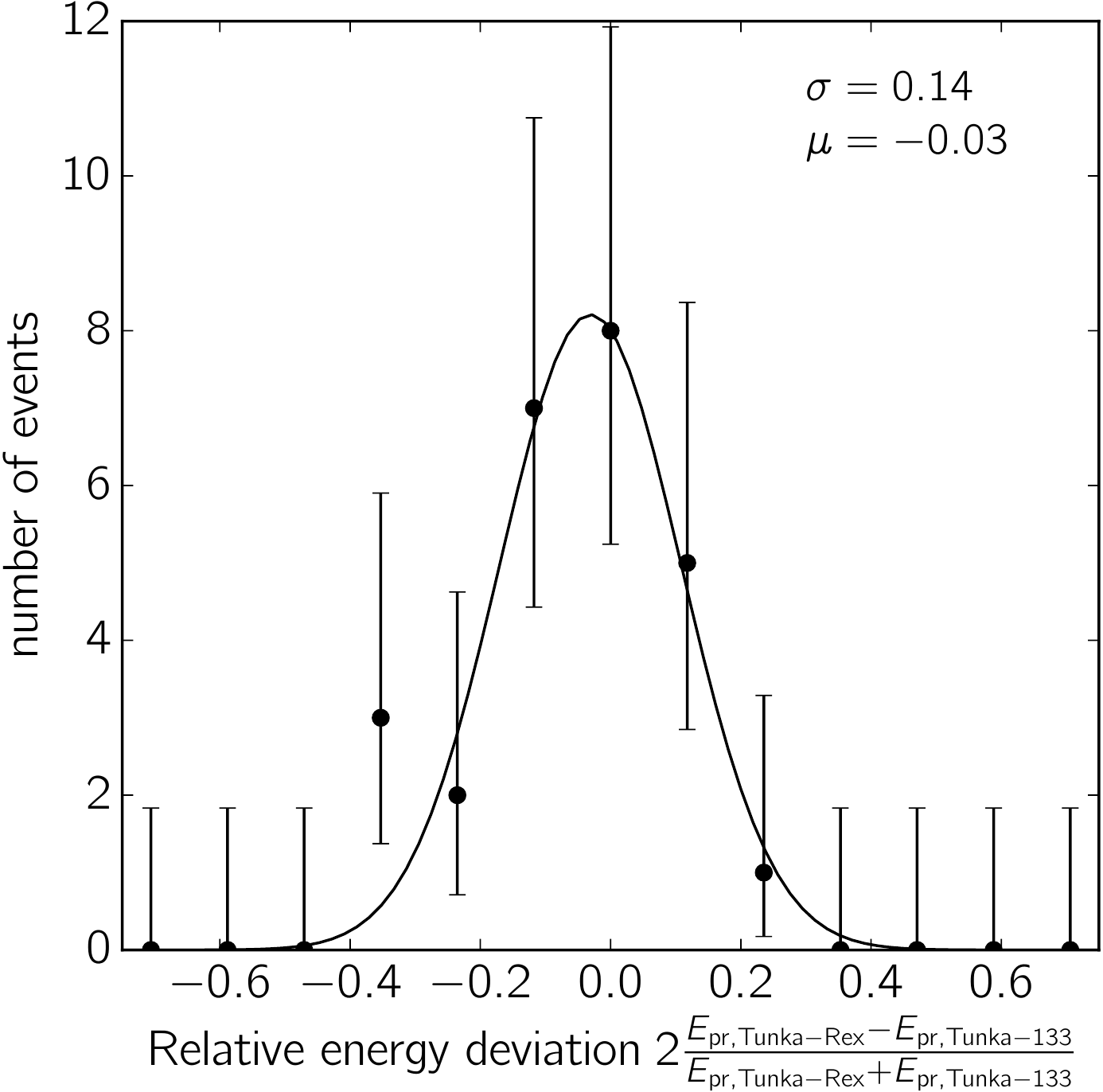}\\
~\\
$10^{17.5}\le E<10^{17.7}$~eV\\
\includegraphics[width=0.49\linewidth]{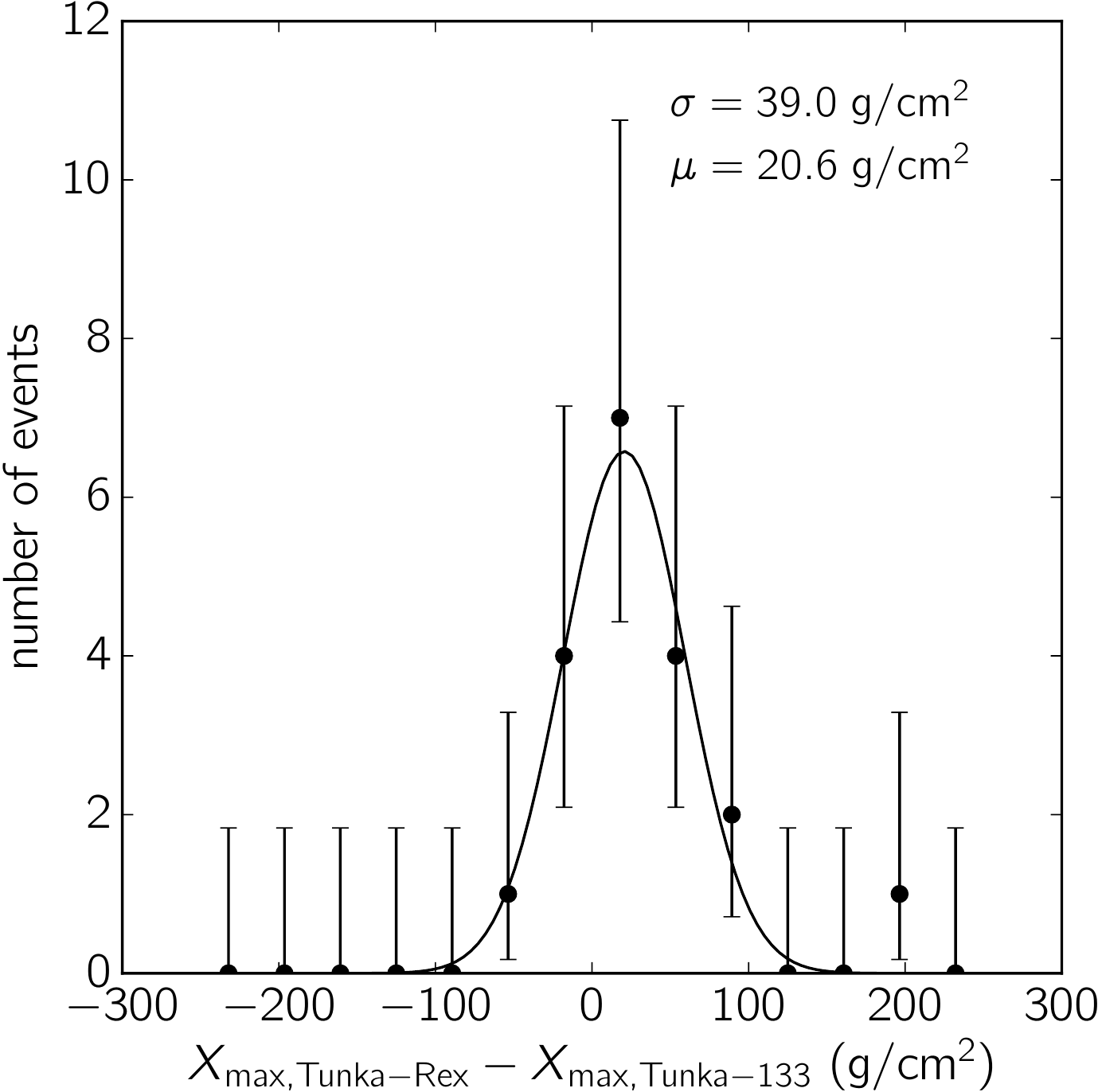}
\includegraphics[width=0.49\linewidth]{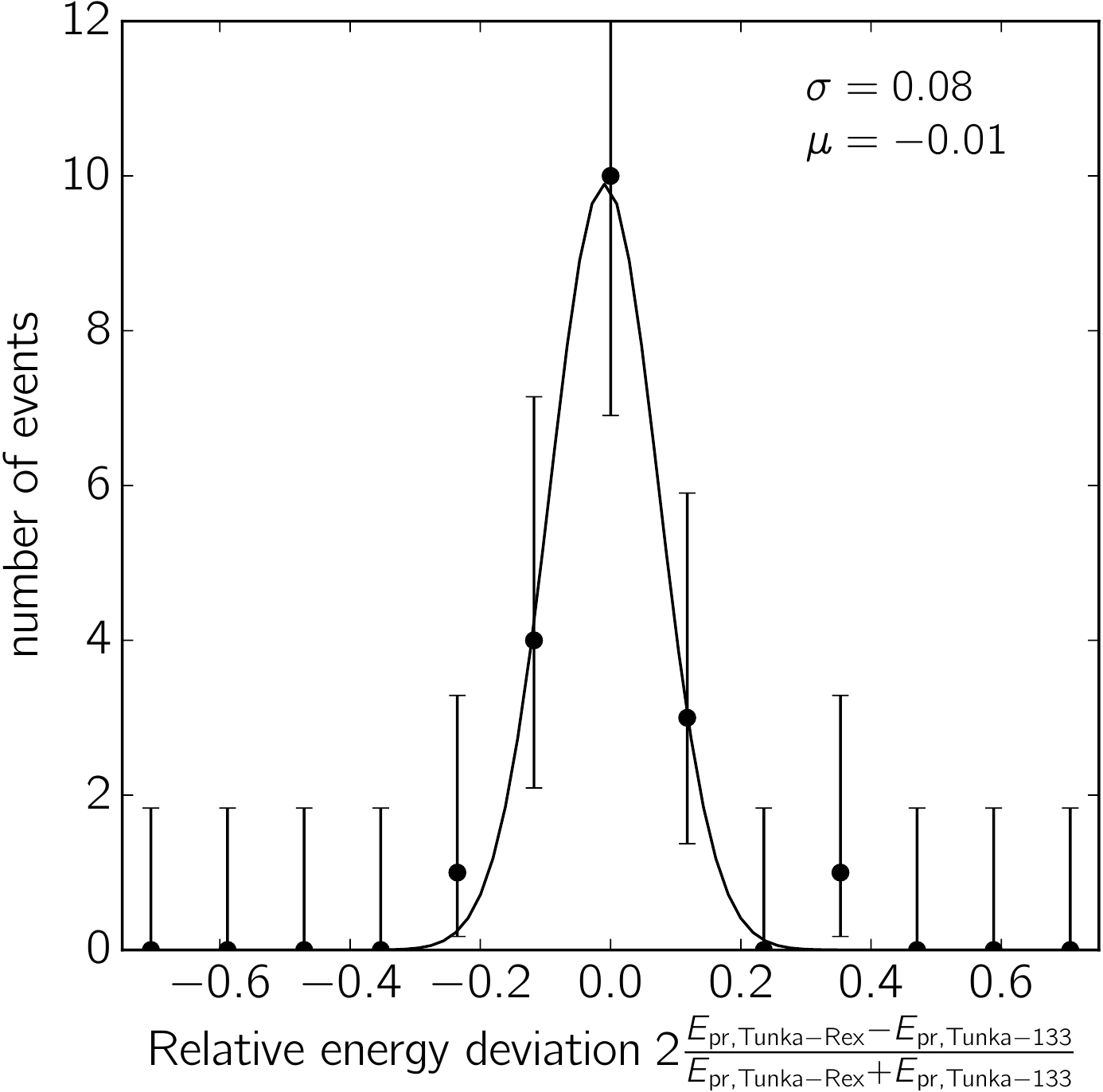}\\
~\\
$E>10^{17.7}$~eV\\
\includegraphics[width=0.49\linewidth]{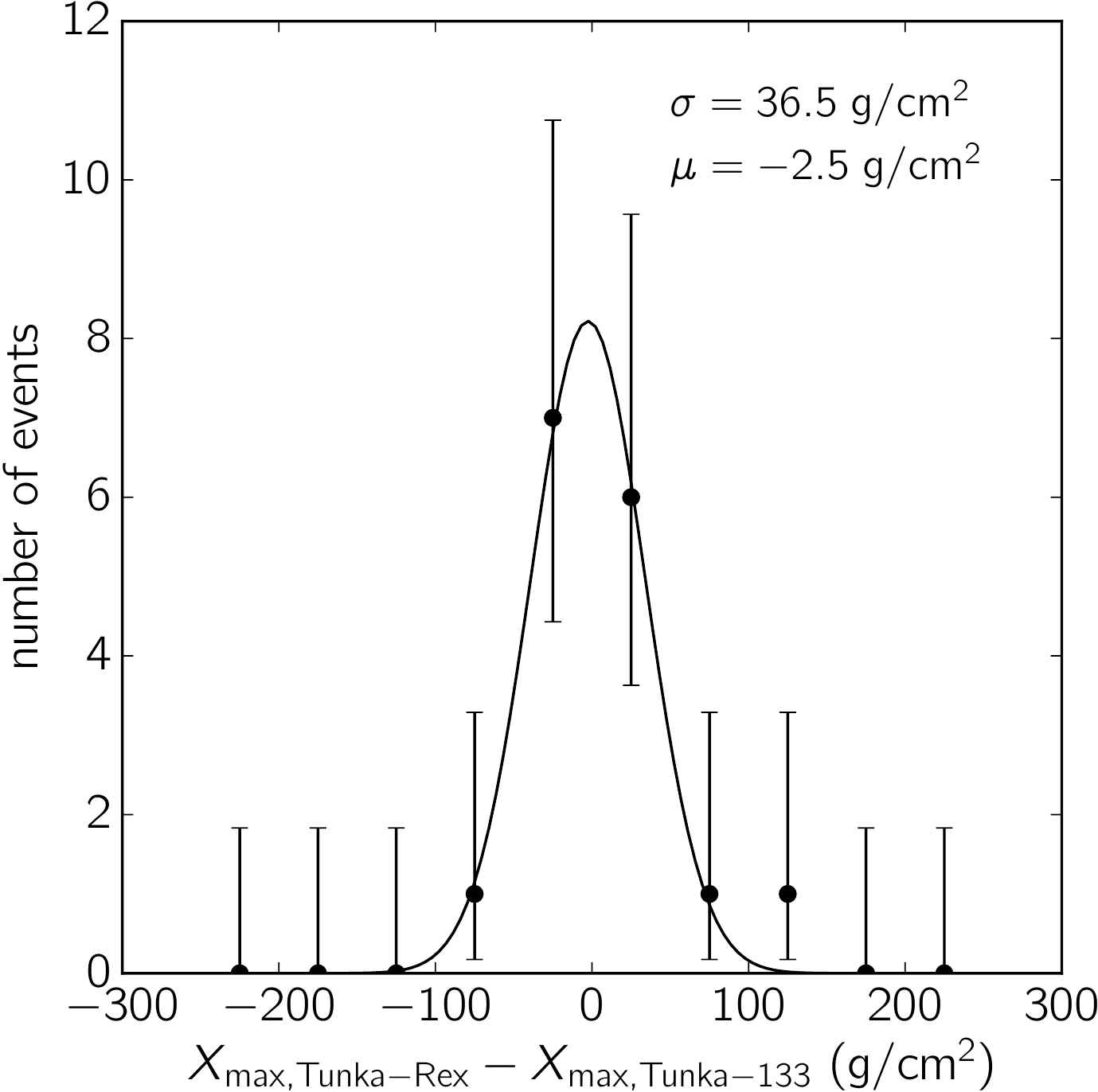}
\includegraphics[width=0.49\linewidth]{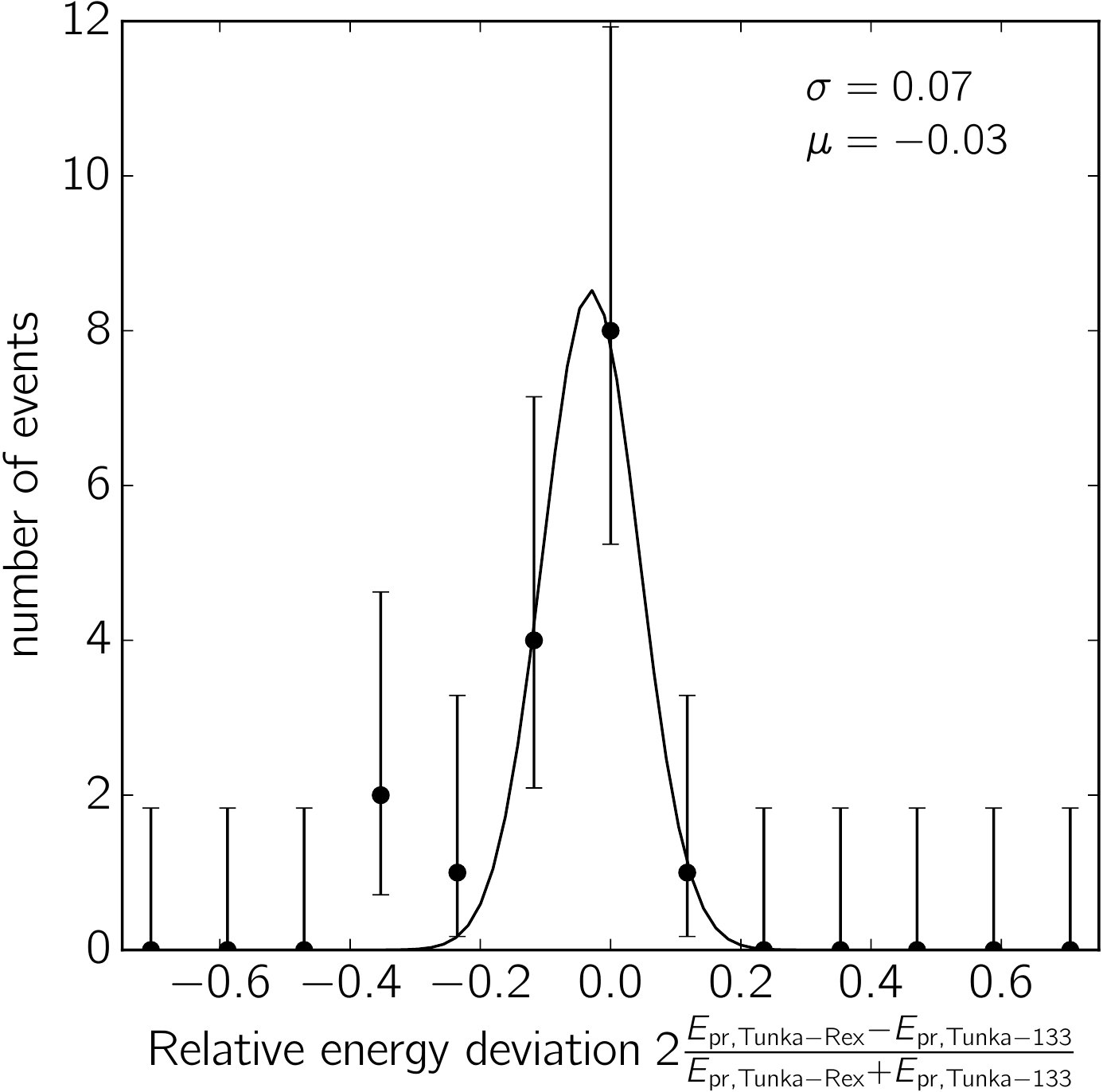}\\
\caption{Relative deviations of shower maximum and primary energy reconstructed by the Tunka-Rex and Tunka-133 experiments for the different energy bins.}
\label{fig:resolutions}
\end{figure}

Here we provide plots (Figs.~\ref{fig:correlations}, \ref{fig:resolutions}) showing the comparison between Tunka-133 and Tunka-Rex as a function of energy.
We divided all data sets into three energy bins (as in Table~\ref{tab:resolution}) with approximately equal numbers of events (about 20) in each bin.
The last season (2016/17) has been omitted from the cross-check due to the lack of calibration of Tunka-133.

\section{Distribution of selected events}
\label{append:distr}
\begin{figure}
\includegraphics[width=0.49\linewidth]{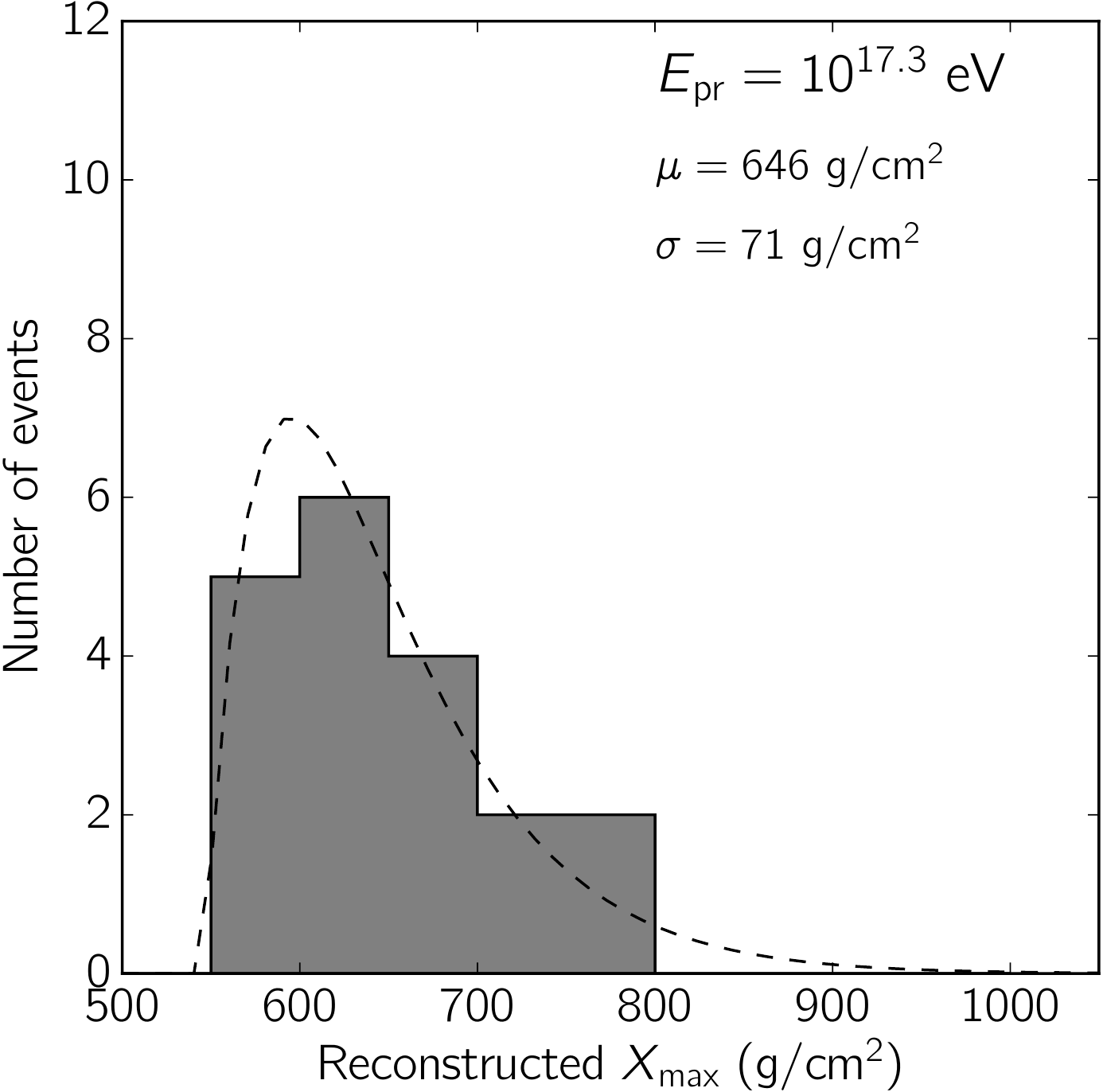}
\includegraphics[width=0.49\linewidth]{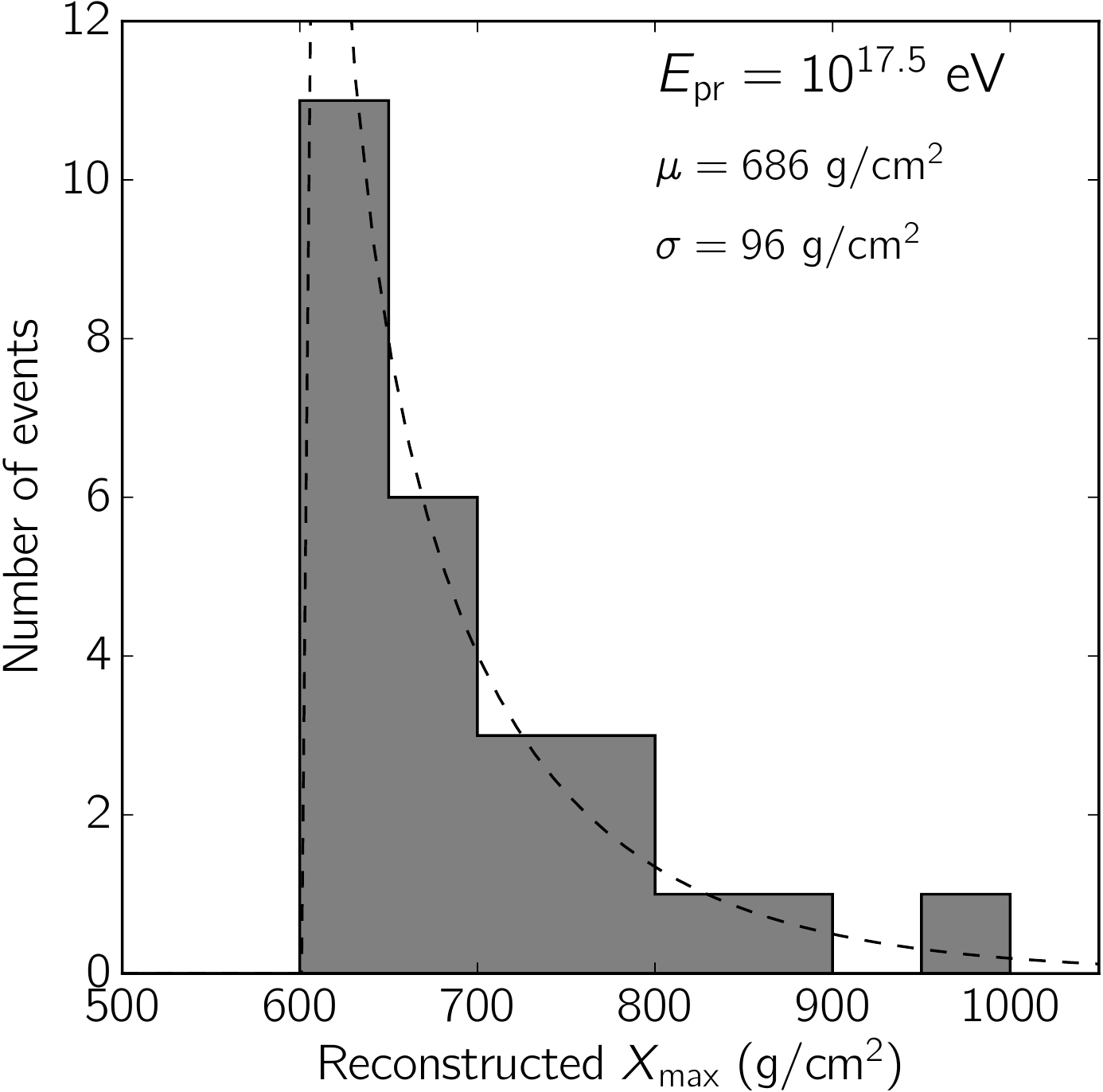}\\
~\\
\includegraphics[width=0.49\linewidth]{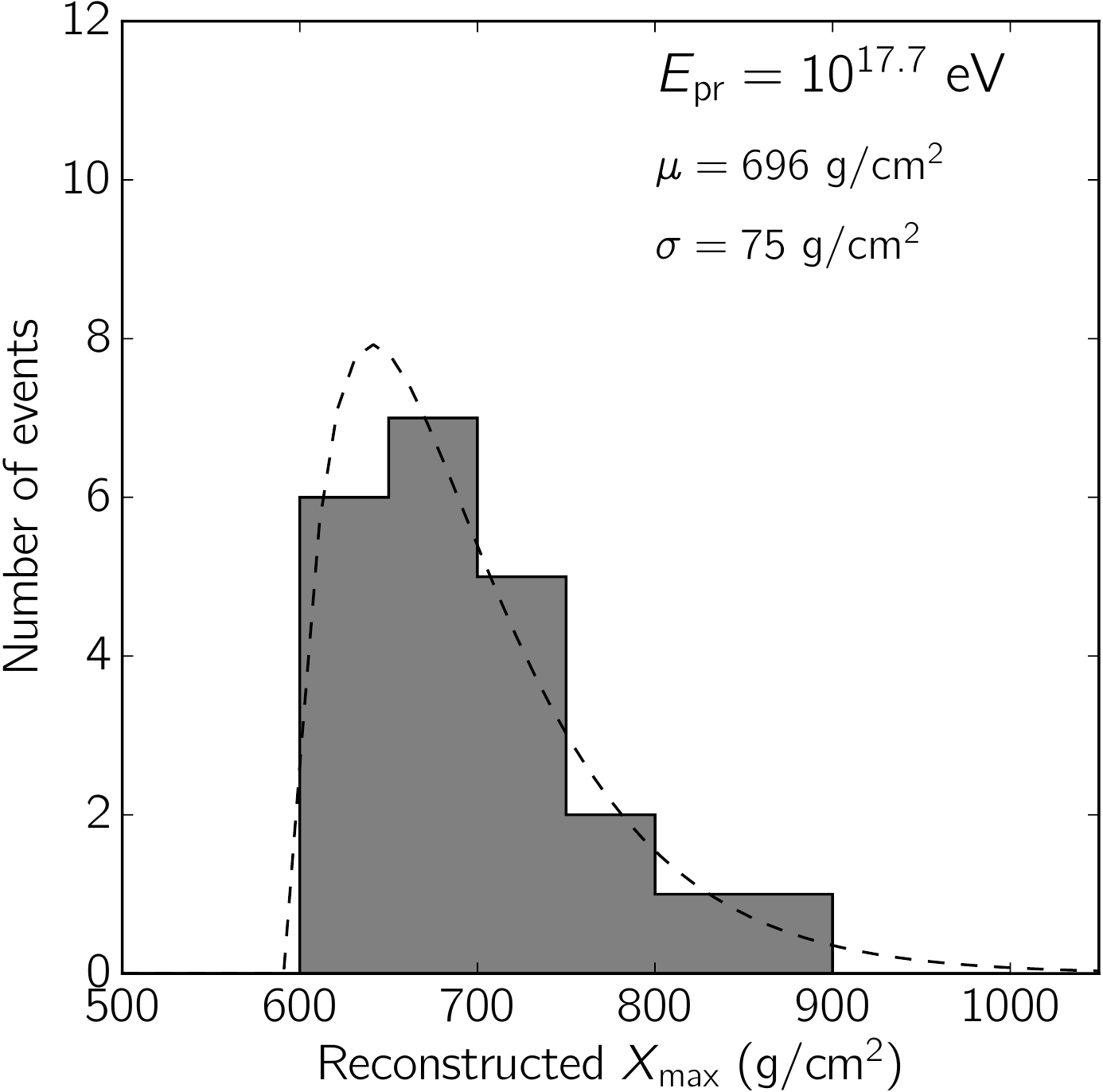}
\includegraphics[width=0.49\linewidth]{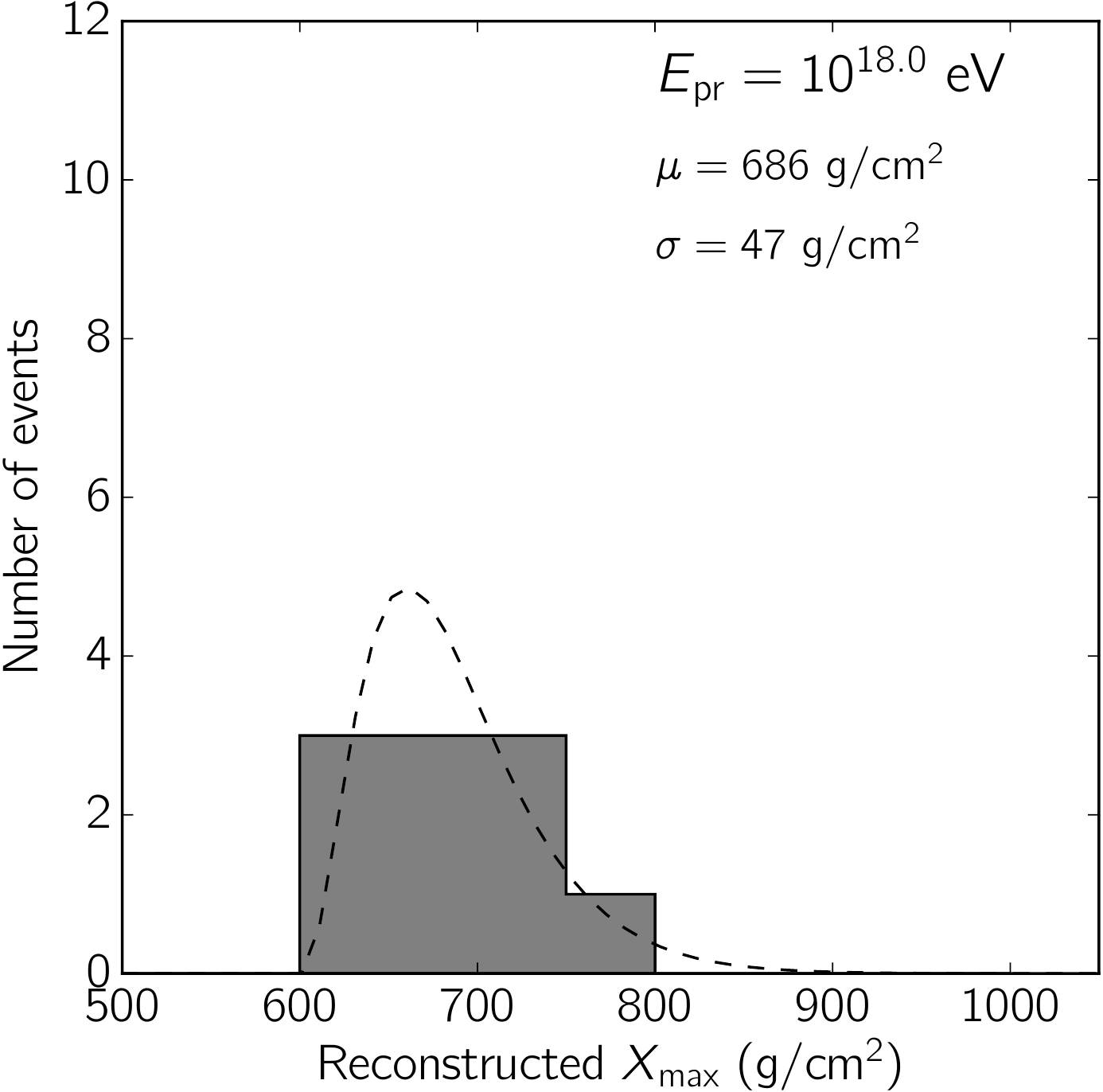}\\
\caption{The distributions of the reconstructed depths of shower maximum
binned by energy.
Since the distribution is non-Gaussian the gamma distribution is chosen to describe the data (dashed line).
The mean values and variations are calculated from the gamma distribution parameters using standard formulas.}
\label{fig:distr_xmax}
\end{figure}

Here we provide the detailed distributions of $X_\mathrm{max}$ presented in Fig.~\ref{fig:xmax_spectrum}.
Since these distributions can be described as convolutions similar to a gamma distribution~\cite{Belov:2006mb}, we fit experimental points with a gamma distribution and derive the mean and standard deviation from the fit.
These values converge to the arithmetic mean and standard deviation (used by other experiments) for large numbers.
The distribution of events as a function of primary energy is given in Fig.~\ref{fig:distr_xmax}.
The values and uncertainties of the Tunka-Rex points are given in Table.~\ref{tab:Xmax_vs_Energy}.

\begin{table}[h!]
\caption{The values and uncertainties of the Tunka-Rex mean $X_\mathrm{max}$ in Fig.~\ref{fig:xmax_spectrum}. 
The uncertainties are calculated as uncertainties of the mean value (standard deviation divided by the square root of the number of events).}
\label{tab:Xmax_vs_Energy}
\begin{ruledtabular}
\begin{tabular}{lllll}
$\log(E_\mathrm{pr}/\mathrm{eV})$ &  \# of events &      $\langle X_\mathrm{max} \rangle$ \\
17.3                              & 19            & $646\pm16$ \\
17.5                              & 26            & $686\pm19$ \\
17.7                              & 22            & $696\pm16$ \\
18.0                              & 10            & $686\pm15$ \\
\end{tabular}
\end{ruledtabular}
\end{table}

\begin{acknowledgments}
The construction of Tunka-Rex was funded by the German Helmholtz association and the Russian Foundation for Basic Research (Grant No. HRJRG-303).
This work has been supported by the Helmholtz Alliance for Astroparticle Physics (HAP),
the Deutsche Forschungsgemeinschaft (DFG Grant No. SCHR 1480/1-1),
the Russian Federation Ministry of Education and Science (Tunka shared core facilities, unique identifier RFMEFI59317X0005,
3.9678.2017/8.9,
3.904.2017/4.6,
3.6787.2017/7.8, 3.6790.2017/7.8),
the Russian Foundation for Basic Research (Grants No. 16-02-00738, No. 17-02-00905) and the Grant No. 15-12-20022 of the Russian Science Foundation (sections~\ref{sec:signalrec} and~\ref{sec:efficiency}).
In preparation of this work we used calculations performed on the HPC-cluster ``Academician V.M. Matrosov''~\cite{HPC_Matrosov} and on the computational resource ForHLR II funded by the Ministry of Science, Research and the Arts Baden-W\"urttemberg and DFG (``Deutsche Forschungsgemeinschaft'').
A part of the data analysis was performed using the radio extension of the Offline framework developed by the Pierre Auger Collaboration~\cite{Abreu:2011fb}.
We thank David Butler for the proofreading of the present paper.
\end{acknowledgments}

\bibliography{references}

\end{document}